\documentclass[useAMS,usenatbib]{mn2e}

\usepackage{graphicx}
\usepackage{color}

\newcommand{\ud}{\,\mathrm{d}}

\title[Large scale structure in GAMA]
	{Galaxy And Mass Assembly (GAMA): The large scale structure of galaxies and comparison to mock universes}

\author[M. Alpaslan et al.]
	     {Mehmet Alpaslan$^{1,2}$, Aaron S.G.~Robotham$^2$, Simon Driver$^{1,2}$, Peder Norberg$^{3}$,\newauthor
	     Ivan Baldry$^4$, Amanda E. Bauer$^5$, Joss Bland-Hawthorn$^6$, Michael Brown$^7$,\newauthor
	     Michelle Cluver$^5$, Matthew Colless$^8$, Caroline Foster$^5$, Andrew Hopkins$^5$, \newauthor
	     Eelco Van Kampen$^9$, Lee Kelvin$^{10}$, Maritza A. Lara-Lopez$^5$, Jochen Liske$^9$ \newauthor
	     Angel R. Lopez-Sanchez$^5$, Jon Loveday$^{11}$, Tamsyn McNaught-Roberts$^3$, \newauthor
	     Alexander Merson$^{3,12}$, Kevin Pimbblet$^{7}$\\
$^1$SUPA, School of Physics and Astronomy, University of St Andrews, North Haugh, St Andrews, Fife, KY16 9SS, UK\\
$^2$International Centre for Radio Astronomy Research, 7 Fairway, The University of Western Australia, Crawley, Perth,\\
 Western Australia 6009, Australia\\
$^3$Institute for Computational Cosmology, Department of Physics, Durham University, South Road, Durham, DH1 3LE, UK\\
$^4$Astrophysics Research Institute, Liverpool John Moores University, IC2, Liverpool Science Park, 146 Brownlow Hill,\\
 Liverpool L3 5RF, UK\\
$^5$Australian Astronomical Observatory, PO Box 915, North Ryde, NSW 1670, Australia\\
$^6$Sydney Institute for Astronomy, University of Sydney, School of Physics A28, NSW 2006, Australia\\
$^7$School of Physics, Monash University, Clayton, Victoria 3800, Australia\\
$^8$Research School of Astronomy and Astrophysics, The Australian National University, Canberra, ACT 2611, Australia\\
$^9$ESO, Karl-Schwarzschild-Str. 2, D-85748 Garching bei München, Germany\\
$^{10}$Institut f\"{u}r Astro- und Teilchenphysik, Universit\"{a}t Innsbruck, Technikerstra{\ss}e 25, 6020 Innsbruck, Austria\\
$^{11}$Astronomy Centre, University of Sussex, Falmer, Brighton, BN1 9QH, UK\\
$^{12}$University College London, Dept. of Physics \& Astronomy, Kathleen Lonsdale Building, Gower Place,\\
 London, WC1E 6BT, UK}

\date{Accepted for publication in MNRAS on 4 November 2013}

\pagerange{\pageref{firstpage}--\pageref{lastpage}}\pubyear{2013}

\begin{document}

\label{firstpage}

\maketitle

\begin{abstract}
From a volume limited sample of 45,542 galaxies and 6,000 groups with $z \leq 0.213$ we use an adapted minimal spanning tree algorithm to identify and classify large scale structures within the Galaxy and Mass Assembly (GAMA) survey. Using galaxy groups, we identify 643 filaments across the three equatorial GAMA fields that span up to 200 $h^{-1}$ Mpc in length, each with an average of 8 groups within them. By analysing galaxies not belonging to groups we identify a secondary population of smaller coherent structures composed entirely of galaxies, dubbed `tendrils' that appear to link filaments together, or penetrate into voids, generally measuring around 10 $h^{-1}$ Mpc in length and containing on average 6 galaxies. Finally we are also able to identify a population of isolated void galaxies. By running this algorithm on GAMA mock galaxy catalogues we compare the characteristics of large scale structure between observed and mock data; finding that mock filaments reproduce observed ones extremely well. This provides a probe of higher order distribution statistics not captured by the popularly used two-point correlation function.
\end{abstract}

\section{Introduction}

Many of the earliest galaxy surveys, such as the CfA Redshift Survey \citep{DeLapparent1986} paved the way in recognising structure in the distribution of galaxies in the Universe. Galaxies tend to cluster into groups, which themselves form the building blocks of large scale structure we observe today \citep{Press1974,Bahcall1988,Bond1996,Eke2004}. The modern view of the cosmic web is that it is composed of clusters and superclusters of galaxies that are connected to each other by groups of galaxies (e.g. \citealp{Bharadwaj2004,Colberg2005,Novikov2006}). These structures themselves surround voids, which are extremely underdense regions containing a small number of isolated galaxies. Therefore, galaxies can be classified as belonging to different types of density regions: filaments, clusters, or voids, with each classification presenting a unique environment to that galaxy.

The evolutionary fate of a galaxy is intimately linked to its neighbourhood; this is well established for scales below 1 Mpc (e.g. \citealp{Hahn2007}). Many observable properties of a galaxy are greatly influenced by the presence of other galaxies nearby; stellar populations in particular are very susceptible to environment. The proximity of galaxies can often trigger dormant regions of gas into infall, leading to an increased rate of star formation \citep{Porter2008}. The local environment of a galaxy has profound effects on many other properties, including colour \citep{Kreckel2012}, stellar mass \citep{Chabrier2003}, gas content \citep{Beygu2013,Benitez-Llambay2013}, luminosity function (\citealp{Croton2005}; McNaught-Roberts et al., in prep) and morphology \citep{Dressler1997}.

Our understanding of large scale structure has developed over recent years, with advanced simulations such as those by \citet{Angulo2012,Habib2012} and large galaxy surveys like the 2dFGRS \citep{Colless2001}, the MGC \citep{Liske2003}, the SDSS-DR7 \citep{Abazajian2009}, the 6dFGS \citep{Jones2009} and GAMA (\citealp{Driver2011}; Liske et al. in prep) progressing side by side. There is still some work to be done, however, on bridging the gap between observations and simulations, in order to establish whether the larger scale environment ($\geq 1h^{-1}$ Mpc) of galaxies influences their evolution. In other words, is a galaxy in a filament discernibly different from a galaxy in a void? If so, how can we use direct observations and simulations to find out? 

Answering these questions requires a robust and reproducible definition of what constitutes a filament and a void. The field of filament finding and classification has been expanding, with numerous algorithms currently being used to detect, classify and link large scale structure to cosmological models \citep{Sahni1998,Pimbblet2005,Forero-Romero2009,Aragon-Calvo2010,Stoica2010,Murphy2011,Sousbie2011,Hoffman2012,Smith2012}. In complement, there is a large volume of work that is currently being done to identify voids in space; regions that are largely underdense compared to the rest of the Universe \citep{El-Ad1997, Peebles2001,Hoyle2004,Thompson2011}. Recently, \citet{Tempel2013} have used a modified marked point process method to search for filaments within a $0.009 \leq z \leq 0.155$ slice of the SDSS, modelling the filamentary network as a series of connected cylinders. Using narrow cylinders (of radius 0.5 $h^{-1}$ Mpc) they identify filaments as having a characteristic length of 60 $h^{-1}$ Mpc, and that galaxies in filaments contribute to 35-40\% the total galaxy luminosity function.

The Galaxy And Mass Assembly (GAMA) survey (\citealp{Driver2009, Driver2011}; Liske et al. in prep) is an ongoing spectroscopic galactic survey aiming to span $\sim 290$ deg$^2$ and to obtain $\sim$ 300,000 galaxy redshifts out to a magnitude of $m_r = 19.8$ mag. A large number of data products for GAMA have already been produced, including catalogues of multi-band matched aperture photometry (\citealp{Hill2011}, Liske et al. in prep), structural analysis \citep{Kelvin2012}, spectral properties \citep{Hopkins2013}, and most importantly for this work, a group catalogue \citep{Robotham2011}.

In this work we introduce an algorithm to identify and classify large scale structures in the three equatorial GAMA fields. We present a series of catalogues that identify different populations of galaxies belonging to distinct types of large scale environments. We are able to detect filaments of groups and galaxies, as well as smaller coherent structures formed by individual galaxies on the peripheries of filaments, dubbed `tendrils', and galaxies that lie in very underdense regions of space, referred to in this work as void galaxies. The aim of this work is to create a structure finding algorithm that is robust, easy to replicate by others, computationally efficient, and mathematically uncomplicated, thereby being as accessible as possible. 

Section 2 introduces the GAMA Group Catalogue and its corresponding mocks, and the sample selection process. In Section 3 we introduce our structure finding algorithm and give an overview of the resulting large scale structure catalogue. Finally, in Section 4 we compare the filaments found in the observed GAMA data to filaments obtained from GAMA mock galaxy catalogues. Throughout this paper, we use a cosmology of $\Omega_{\mathrm{m}} = 0.25,\; \Omega_{\Lambda}=0.75,\; H_0 = h\; 100 \mathrm{km s}^{-1}\;\mathrm{Mpc}^{-1}$, consistent with the cosmology used to create the GAMA mocks (as described in \citet{Robotham2011,Merson2013}).

\section{Data}

\subsection{GAMA Group Catalogue}

The GAMA survey currently spans across three equatorial fields measuring $12 \times 5$ deg$^2$ centred at $\alpha$ = 9h, $\delta$ = 0.5 deg (G09), $\alpha$ = 12h, $\delta$ = -0.5 deg (G12) and $\alpha$ = 14.5h, $\delta$ = -0.5 deg (G15), out to $m_r = 19.8$ mag and two southern fields at $\alpha$ = 02h, $\delta$ = -7 deg (G02) and $\alpha$ = 23h, $\delta$ = -32.5 deg.

One of the major data products of GAMA is the GAMA Group Catalogue (\citealt{Robotham2011}; hereafter R11), providing a comprehensive catalogue of 23838 galaxy groups across the three equatorial GAMA fields out to $m_r <  19.8$ mag in the three equatorial GAMA regions. Note that we use the GroupFindingv06 catalogue, which is an updated version of the catalogue presented in R11, containing more objects. Whenever we refer to results from R11, we refer to this updated catalogue. The final catalogue contains 73298 galaxies out of a possible 180979, roughly 60\% of all galaxies. Notably, most groups found in the catalogue are galaxy-galaxy pairs that span across the entire redshift range. GAMA is a highly complete spectroscoic survey ($~98\%$ as of the creation of the group catalogue, with measured redshifts having an uncertainty $\sigma_v \approx 50$ kms$^{-1}$). The average target density is 1050 galaxies per square degree, out to $m_r < 19.8$ mag. This means that galaxies that may previously have been considered to be in the field are now seen to be part of an underlying group of faint galaxies (R11). Similarly, regions thought to contain few galaxies are now seen to contain not just more galaxies, but a considerable amount of structure. This is one of the principal strengths of GAMA, and is fundamental to why it is so well suited for studies of structure. 

\begin{figure*}
	\includegraphics[width=0.9\textwidth]{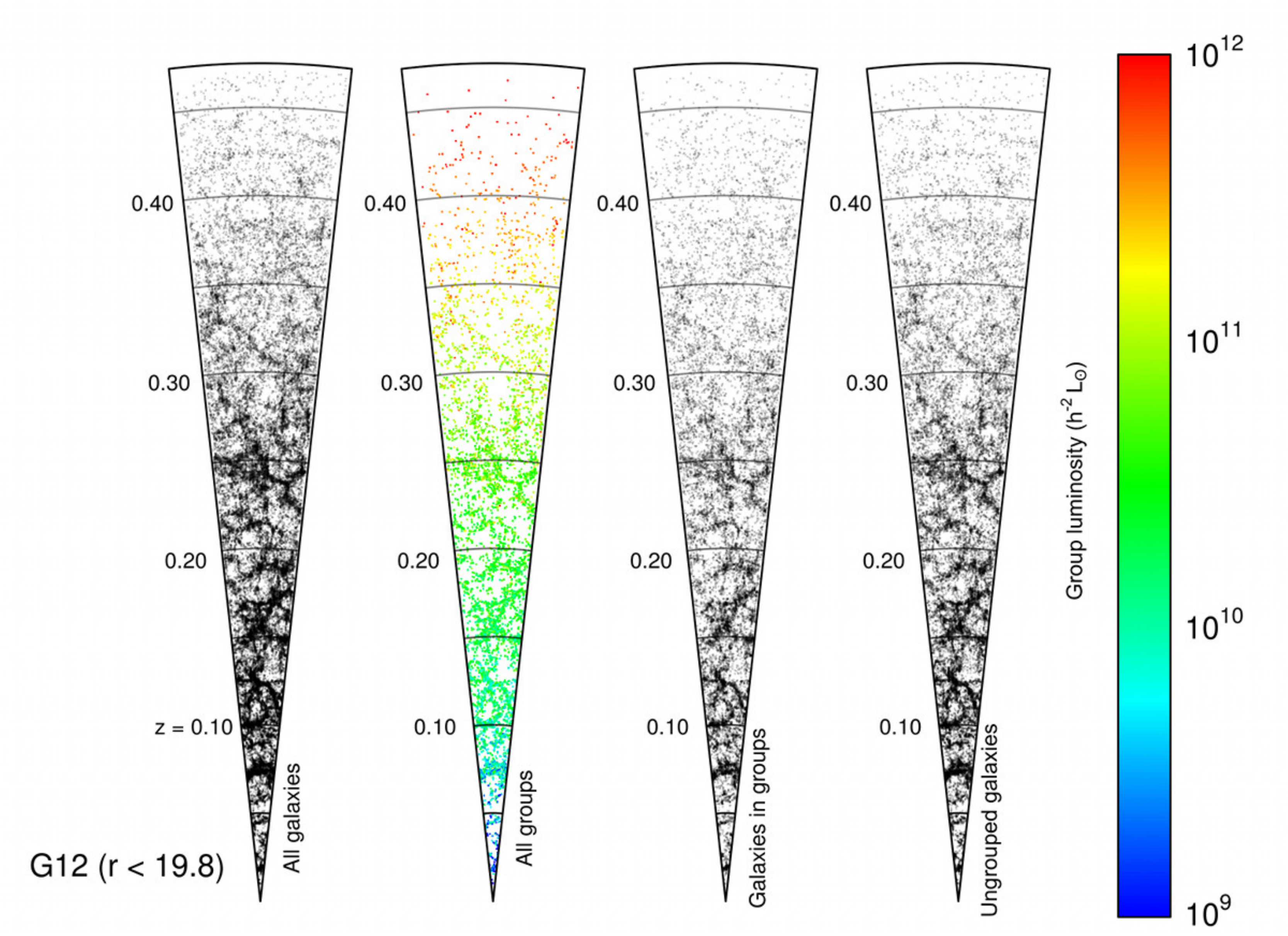}
	\caption{Side by side comparison of different structures in the G12 region of GAMA. From left to right, the cones display all galaxies with $m_r < 19.8$, groups of galaxies as identified in R11 and coloured by their total group luminosity, all galaxies that belong to these groups, and the remaining, ungrouped galaxies. Here we define any galaxy that is not in a group as being an isolated galaxy. Isolated galaxies continue to trace large scale structure and must be considered when searching for filaments.}
	\label{fig:g12_194}
\end{figure*}

At the heart of the process used to generate the GAMA Galaxy Group Catalogue (G$^3C$) is a friends-of-friends algorithm that operates on projected and radial separations independently (see Figure 1 in R11). This is a very important step, as it allows the algorithm to take redshift space distortions into account. Figure \ref{fig:g12_194} displays four panels with different populations of galaxies and groups for the G12 region. The two panels to the left show, respectively, all galaxies that are within this region, and all groups recovered by the FoF group finder (coloured by their group luminosity in $L_{\odot} h^{-2}$). The third panel shows all galaxies in groups. The final panel shows all galaxies not in groups, which we define as being isolated galaxies. This final population is very important, as it highlights features of large scale structure that are not characterised by metrics that rely on local overdensities of galaxies. This `tendril' population emphasises that large-scale structure exists on all scales, persisting even down to rather low values of local galaxy density. A complete quantification of large-scale structure must, therefore, not rely solely on a threshold in density, but must take into consideration the spatial distribution of galaxies themselves. However, it is important to note that some of these field galaxies will belong to undetected low mass groups. Throughout this paper, we define any galaxy that is not in a group as being an isolated galaxy.

The G$^3$C provides estimates of group centres, a number of size estimates, integrated magnitude and luminosity measurements, and other properties for each group. Of greatest importance to our work are the position estimates for each group, as the groups provide the first step in generating filamentary structure. The projected group centre is defined by determining the $r_{AB}$-band luminosity of each galaxy in the group and calculating the centre of light (CoL), then iteratively discarding the galaxy furthest from the CoL until two galaxies remain, at which point the brightest $r_{AB}$-band galaxy is chosen as the group centre.

\subsection{GAMA mocks}

The GAMA mock galaxy catalogues (R11; \citealp{Merson2013}) are 9 lightcones that match the geometry of the three equatorial GAMA fields. They are built by populating dark matter haloes within the Millennium Simulation \citep{Springel2005} using the GALFORM semi-analytic galaxy formation model \citep{Cole2000}, following the \citep{Bower2006} description. The free parameters in the Bower06 model were tuned to approximatively reproduce the local galaxy and stellar mass function using data available at that time. Since 2006, more detailed measurements have been obtained and to provide an exact match to the GAMA survey and hence an identical selection function, the lightcone luminosity functions are abundance matched to the GAMA luminosity function \citep{Loveday2012} in the \emph{r}-band. The abundance matching results in small magnitude changes (typically less than 0.1) of the original GALFORM model predictions. Such changes are in line with the expected difference between different magnitude definitions, which are not included in GALFORM.

The mock catalogues purpose within the context of the group catalogue is to provide a set of galaxies whose true grouping is known. They can then be used to optimise the various parameters of the FoF algorithm used in R11. In other words, the grouping should be bijectively matched. Mathematically, bijection refers to a function that provides exact pairings between two sets of elements. In the context of group finding, bijection is used to determine which galaxies from a group recovered by the friends-of-friends algorithm are actually members of an intrinsic group. We consider a match to be bijective if for a given group recovered from the mock galaxy population, at least 50\% of its galaxies must belong to the actual halo they originate from and vice-versa. As each group cannot bijectively match more than one known group, this ensures that there is no ambiguity in the final group catalogue. A second measure of grouping quality relates to how significant the matching between recovered FoF groups and intrinsic groups is. This is defined as the product of the relative fractions of members that belong to the recovered FoF group and the intrinsic group. Reduced to its simplest form, this means a minimum of $1/2 \times 1/2 = 0.25$  matching fraction is required for a bijectively matched group; that is to say, at least 50\% of the members of both groups must belong to the correct group. These measures are condensed into some efficiency statistics, which must be maximised in order to obtain the truest grouping. By generating a large variety of group catalogues on the mocks using different parameters for the FoF algorithm (detailed in R11), it is possible to optimise for the best possible grouping. 

\subsection{Sample selection}

The G$^3$C and its accompanying mock group catalogue, as well as the observed galaxy and mock galaxy catalogues, form the input data sets from which we detect and classify large scale structure. As with any other body of observed data, it is important to ensure that the subsample of galaxies and groups we utilise are as free as possible from any intrinsic bias, most often caused by observational effects and the necessary limitations found in any galaxy survey. GAMA benefits from an exceptionally high spectroscopic completeness ($> 98\%$ for the sample used in defining the group catalogue), so completeness effects are accounted for easily.

We wish to ensure that for a given sample of galaxies and groups, we are observing every possible galaxy (and therefore group) within that absolute magnitude limit, with a fainter limit resulting in more galaxies. This is particularly important for a study on large scale structure, where different populations of galaxies in varying density environments span many magnitude ranges \citep{Driver2011}; it also allows any linear structure finder to use a constant search length instead of varying it as a function of redshift. One must therefore select a luminosity limit that maximises the number of galaxies that are retained after the cut is applied. Within the G$^3$C, the proxy for absolute magnitude for a group is given by the \texttt{TotFluxProxy} parameter. This is defined as the total luminosity for the group, and is corrected to account for selection effects and missing flux, and is given in units of $L_*$.

We can approach this sample selection problem from the other end, and determine the faintest possible galaxy that is visible in GAMA at a given redshift $z$, given our apparent magnitude limit of $m_r  < 19.8$ mag. We calculate the distance modulus of an object at $z$, using the cosmological luminosity distance $D_L$ to that object at that redshift. In other words,

\begin{equation}
DM = 5 \log D_L + 25
\end{equation}

\noindent with $D_L = (1+z)R_0 S_k(r)$, where $R_0S_k(r)$ refers to the radial comoving distance, $D_M$, all given in $h^{-1}$ Mpc. We can then use this to calculate the absolute magnitude of an object with $m_r = 19.8$ mag at redshift $z$, using the $k$-correction taken from R11.



For a given redshift, we calculate the absolute magnitude of the faintest possible galaxy that can be seen within the GAMA survey, given by $M_r^h(z) = 19.8 - DM(z) - k(z)$. We pick a redshift $z$ and discard all galaxies whose magnitude $M_{\mathrm{gal}} > M_r^h(z)$. We then go through the group catalogue and discard any groups that have fewer than 2 members remaining, retaining only groups that would still have been detected with this absolute magnitude cut. We pick $z$ such that we retain the largest number of groups and galaxies. This value is $z = 0.213$, where $M_r^h = -19.77$ mag.

\begin{figure*}
	\centering
	\includegraphics[width=0.45\textwidth]{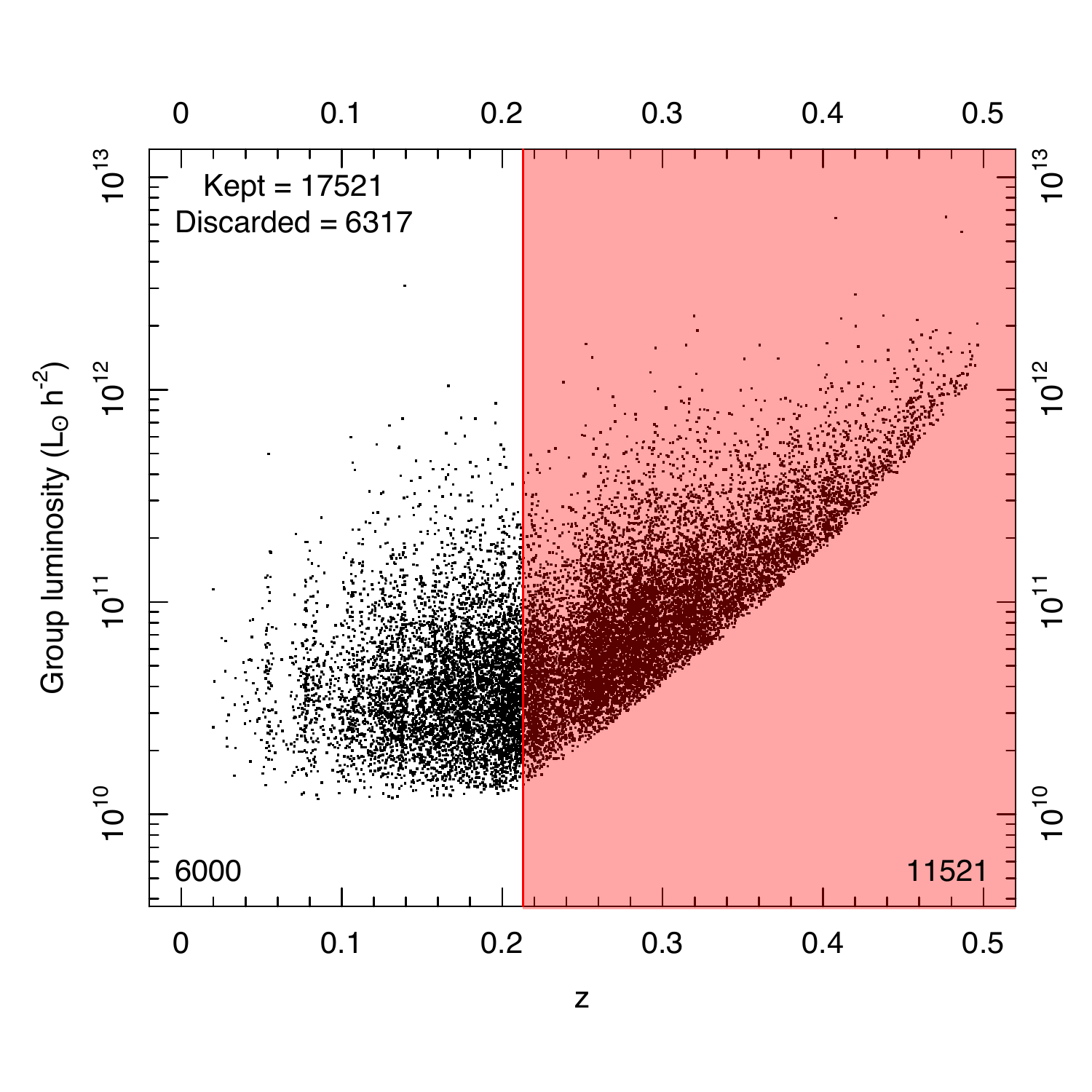} \includegraphics[width=0.45\textwidth]{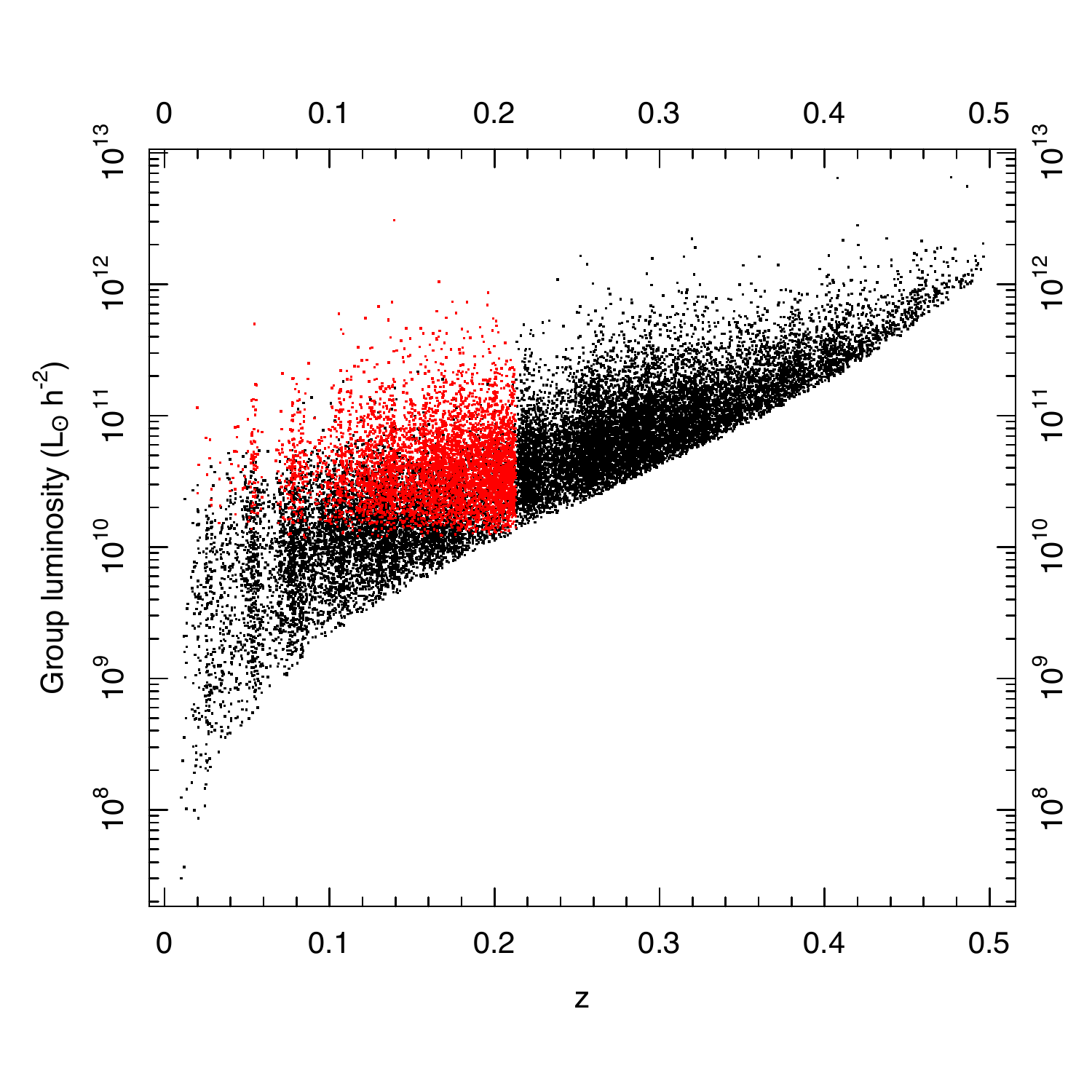}
	\caption{\textit{Left panel:} Distribution of the total group r-band luminosity (see R11 for more details) as a function of redshift, after the sample selection process has been applied. The numbers on the top left display the number of groups kept and discarded after removing galaxies - the kept galaxies are plotted in the figure. The region shaded in highlights the region with $z > 0.213$ and is no longer volume limited, and the numbers in the bottom corners show how many groups are above and below the redshift cut. We are therefore left with 6000 groups across all three GAMA regions, with $z \leq 0.213$ and with at least two or more galaxies with $M_r \leq -19.77$. This sample selection ensures the structures we detect are volume limited. \textit{Right panel:} All groups in the G$^3$C are plotted here, with our final sample shown in red. The redshift limit of $z = 0.213$ is easily seen here. The red sample corresponds to all groups in the unshaded region in the left.}
	\label{fig:z0213}
\end{figure*}

The sample selected by $z = 0.213$ and $M_r^h = -19.77$ mag can be seen in the left panel of Figure \ref{fig:z0213}. The numbers on the top left of this panel refer to the number of groups that are kept after the absolute magnitude cut is applied, and those that are discarded. Only groups that are kept are plotted. The numbers in the bottom left of this plot show the number of galaxies below the redshift limit, and those above (shown in the region shaded in red). The notable feature of this plot is that the group luminosity distribution is effectively flat below the redshift cut. The final sample contains a total of 45542 galaxies and 6000 groups across the three equatorial GAMA regions. 

The right hand panel in Figure \ref{fig:z0213} highlights our sample within the context of the entire G$^3$C, with the group luminosity plotted as a function of redshift for all groups. The points in red show all the groups in our sample, which have at least two members left. For any group left in the sample, we use its full group properties as listed in the G$^3$C. We also apply this same sample selection to the mock galaxy and group catalogues.

\begin{figure*}
	\centering
	\includegraphics[width=0.85\textwidth]{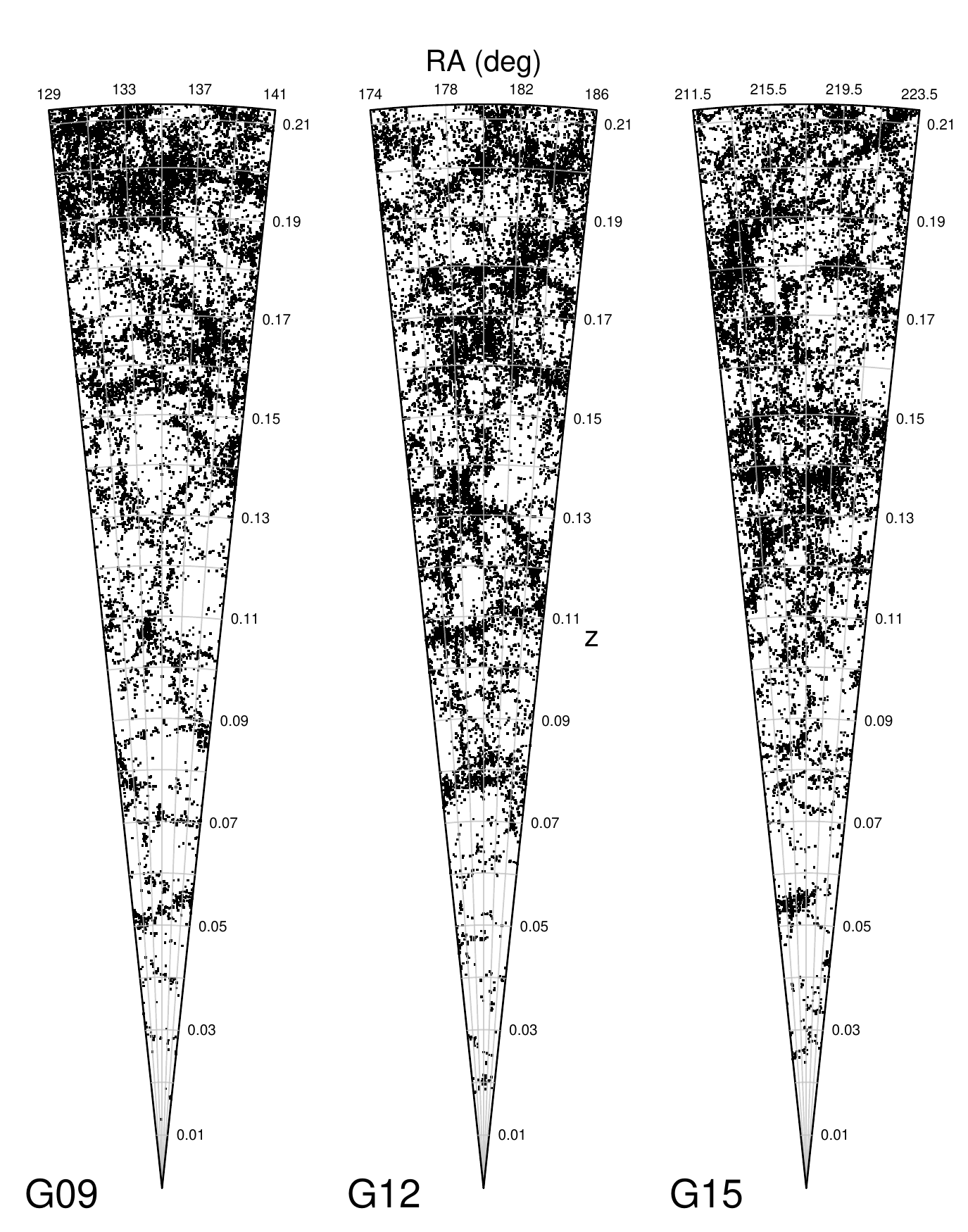}
	\caption{Three side by side cones showing the remaining galaxy sample after the selection described in Figure \ref{fig:z0213} for the G09, G12 and G15 regions respectively out to $z = 0.213$. All three cones span the full $5^{\circ}$ declination range, which results in increasing projection effects at higher redshifts.}
	\label{fig:samplecones}
\end{figure*}

The three panels in Figure \ref{fig:samplecones} show, for the three equatorial fields from G09 to G15, all of the galaxies (grouped and ungrouped) in the selected sample. In all three regions the number density of galaxies increases sharply after $z \approx 0.1$. Lowering the absolute magnitude limit for the sample, thereby selecting more faint galaxies, would reveal more faint galaxies at low redshifts; however we would suffer from a much smaller sample size.

Some of the filaments we detect will be truncated by the survey's edge; or their morphology will lose definition as the algorithm reaches the survey edge. Any kind of distance cut of our sample from the survey edge would produce similar results; neither is it feasible to impose periodic boundaries on our data. We therefore compromise and flag all galaxies and groups that are within 4 $h^{-1}$ Mpc from the survey edge. This value is chosen in accordance with the parameters used to select galaxies near filaments, as described in the next section.

\section{Filaments and large scale structure}

\subsection{Minimal spanning trees and Scooper}

Having selected an appropriate sample, we move to the task of classifying galaxies and groups as being part of the large scale structure of the Universe, or within less dense regions and/or voids. This classification method must, primarily, be easily repeatable and be as objective as possible with regards to classifying large scale structure. Our algorithm works on the basis of two assumptions: (1) that all bright, high luminosity groups tend to live in knots of filaments and that (2) void galaxies are only clustered at extremely small scales. The application of these assumptions is discussed below. 

The filament finder is based on a minimal spanning tree method \citep{BarrowJ.D.1985} and has been used previously by others \citep{Graham1995,Doroshkevich2004,Colberg2007} to examine the large scale structure of galaxies and haloes. Here we apply the minimal spanning tree approach to groups of galaxies, instead of individual galaxies, and build upon its results. This approach of using groups instead of galaxies as tracers of filaments is very similar to the approach used in \citet{Murphy2011}. Minimal spanning trees (MST, \citealp{Iyanaga1980}) are a product of graph theory and are commonly used in a number of scientific fields, including computer science, sociology, scientometrics and epidemiology. They are particularly useful for picking out `skeletal' patterns and linear associations within point data sets and for distinguishing clustering and structure in a systematic and quantitative way, making them ideal tools to objectively detect large scale structure in the Universe.

Within graph theory, a \textit{graph} is a collection of \textit{nodes} (in this case, groups) and \textit{edges} (straight lines connecting nodes). A \textit{path} is defined as a sequence of edges that joins nodes, and a graph where a path is possible between any pairs of nodes is a \textit{connected} graph. A \textit{spanning tree} is defined as a graph where a single path connects all nodes and has no loops. If this path is the shortest possible path that connects all nodes, then it is a \textit{minimal spanning tree} (MST). MST-based algorithms are analogous to FoF-based ones, as an MST is simply one specific solution of a FoF algorithm.

For a selected sample of groups and galaxies, the large scale structure algorithm is composed of 5 main steps:

\begin{enumerate}
\item Generate an MST on group centres, and remove excessively lengthy edges (see Section 3.1.1). The structures that are left over are defined as \textit{filaments}; in other words all groups that are in the same set of unbroken links, or `network' are considered to be part of the same structure.
\item Examine the morphology of each filament by subdividing it into a series of branches, including the \emph{backbone}, which is the longest link that travels from one end of the filament to the other through its most central node.
\item Travel along each filament, scooping up galaxies that lie within a certain orthogonal distance $r$ from each filament. These are referred to as \textit{galaxies near filaments}.
\item Having removed galaxies near filaments from our sample, we generate and trim another MST on these unassociated galaxies. These structures are defined as \textit{tendrils}, containing \textit{tendril galaxies}; as with filaments, all galaxies that belong to the same unbroken chain are considered to be part of the same tendril.
\item Any galaxies not in tendrils or near filaments are finally classified as being \textit{void galaxies}.
\end{enumerate}

\begin{figure*}
	\centering
	\includegraphics[width=1\textwidth]{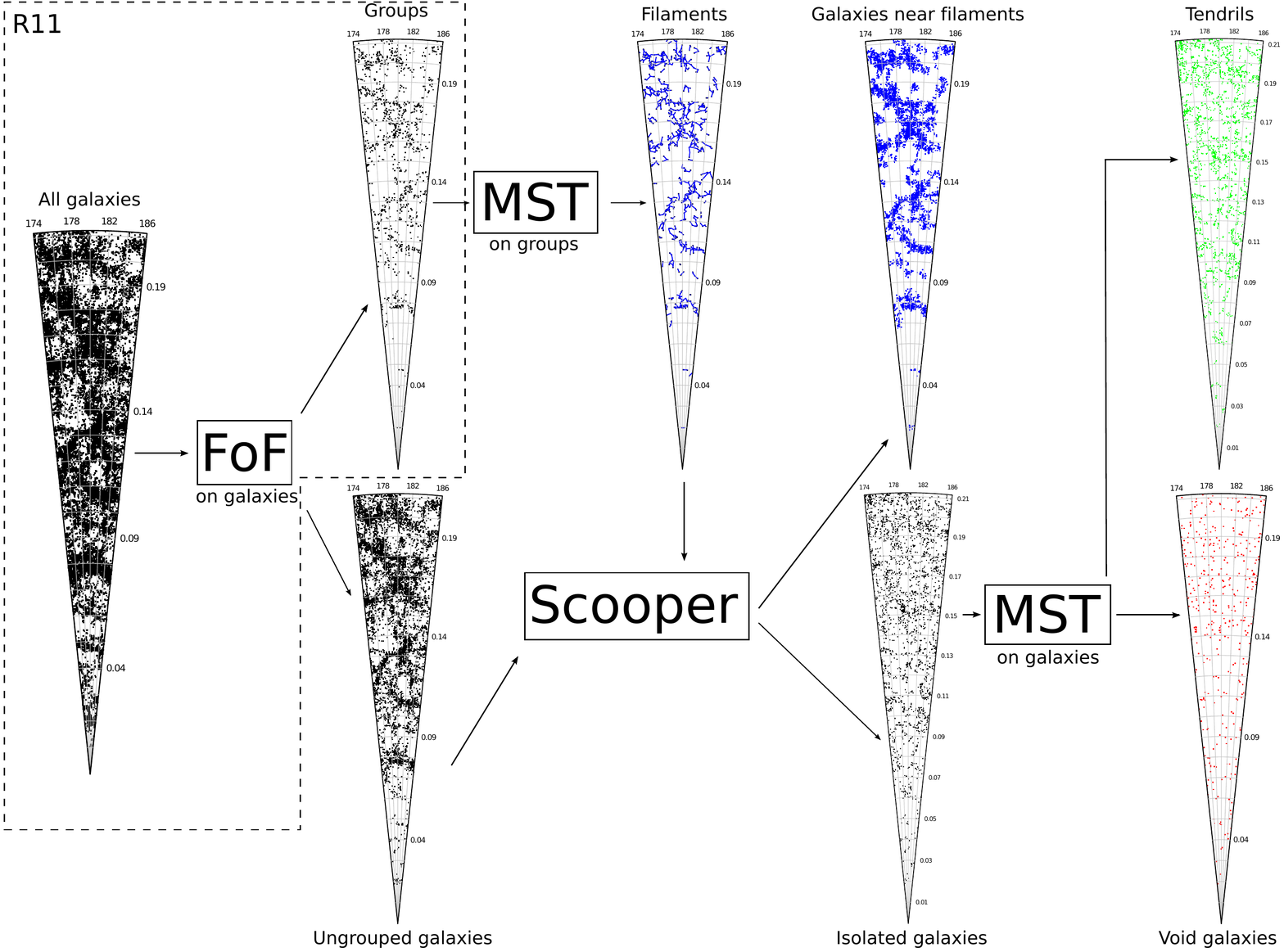}
	\caption{Flowchart schematically describing, for one example region, all the steps taken to go from a distribution of galaxies to a network of filaments, tendrils and voids. We show all groups on the top panel, and all ungrouped galaxies on the lower panel. The groups are then put into a minimal spanning tree and the longest edges are trimmed. Ungrouped galaxies are then scooped up around each filament, giving the large network of galaxies near filaments (shown in blue). All ungrouped galaxies are then classified as being tendril galaxies (in green) or void galaxies (shown in red).}
	\label{fig:flowchart}
\end{figure*}

A visual representation of this algorithm is shown in Figure \ref{fig:flowchart}, where we show the data and output of R11 in the region enclosed by the dashed black lines. The groups are then put through an MST and placed into filaments; which are then combined with galaxies using the \texttt{Scooper} algorithm to identify galaxies near filaments (shown in blue) and isolated galaxies. Isolated galaxies are then put through another MST and classified into tendrils (shown in green) and voids (shown in red). The algorithm then outputs a series of interlinked catalogues that give summary statistics for each filament and tendril, and the associated groups and galaxies within filaments, tendrils and voids through a series of unique identifiers. We now describe the steps given above in greater detail.

\subsubsection{Minimal spanning tree on groups, and filaments}

The construction of the MST on the groups (and, subsequently, the galaxies) is done using the \texttt{nnclust} package within the R programming language. The function \texttt{mst} within \texttt{nnclust} constructs a minimal spanning tree for a set of points on a 2D or 3D Cartesian space using Prim's algorithm \citep{Prim1957}. Prim's algorithm functions on the basis of knowing the distance between all nodes in a graph. Starting from a random node, the algorithm travels along an edge to the nearest node. It then travels to the node nearest to either of the nodes it has already visited, and continues this process iteratively until all nodes have been visited. The path it has taken to do this is the minimal spanning tree.

Comoving Euclidean coordinates of group centres are fed into \texttt{mst}, whose output is a set of links between nodes, and their distances. The links are given by ID names between the start and end of an edge. We reject any edges whose length is beyond a certain threshold value. This allows us to identify distinct sub-structures and removes unrealistically long links between objects in low density regions. Objects that remain in unbroken chains are then grouped together as an individual filament.

\begin{figure*}
	\centering
	\includegraphics[width=0.85\textwidth]{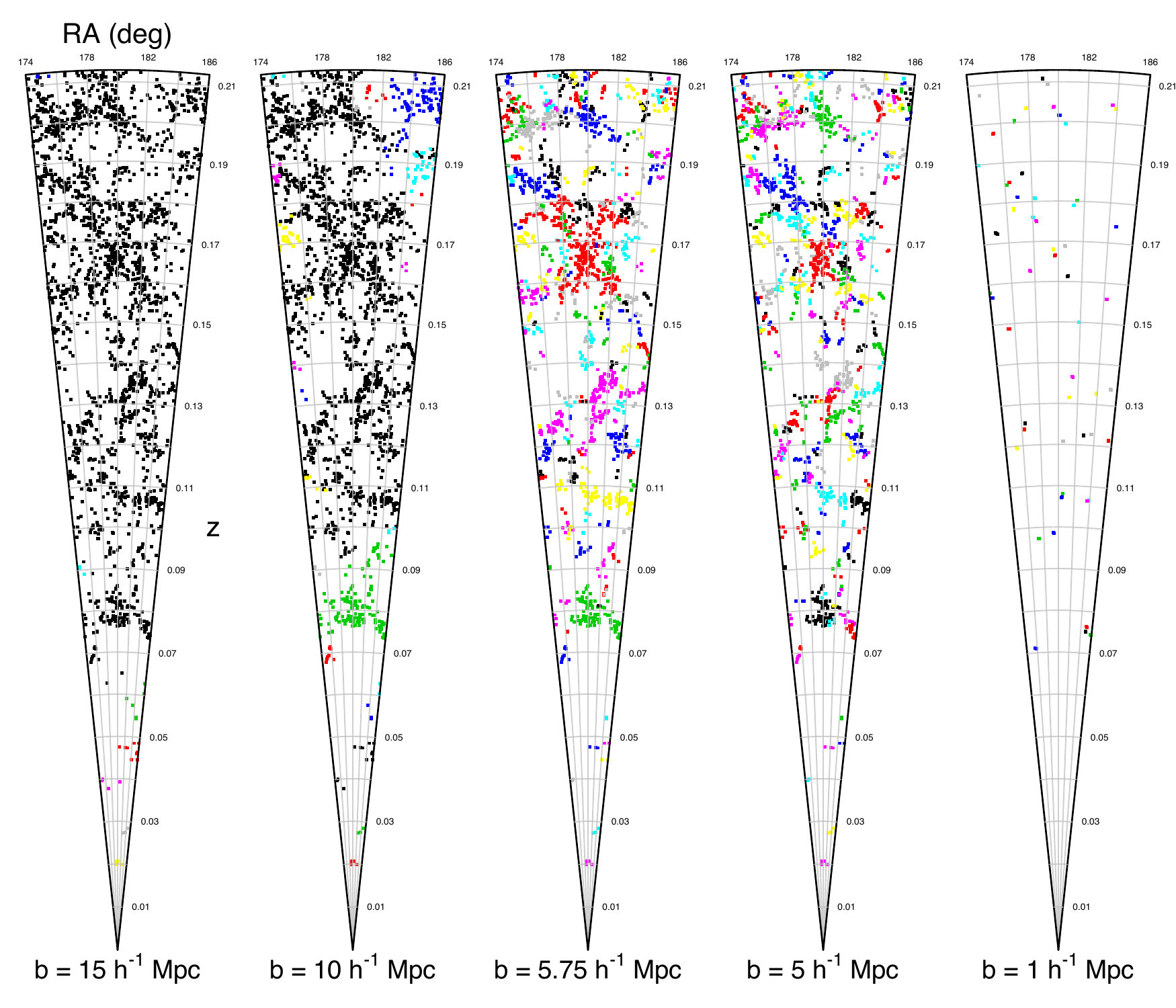}
	\caption{Filaments constructed from the same minimal spanning tree, but with different maximum edge lengths, decreasing from left to right. Groups in the same filament are coloured in matching colours. As the linking length increases, all galaxies tend towards being in one single huge filament, while as it decreases, we are only left with groups that are in close proximity to each other. The number of groups in filaments also drops as $b$ decreases, as groups with no links to other groups are not considered to be filaments (a filament needs at least 2 groups).}
	\label{fig:difftrees}
\end{figure*}

The choice of the maximum edge length $b$, is a vital one. Examples of different maximum linking lengths are shown in Figure \ref{fig:difftrees} where $b$ is given in units of $h^{-1}$ Mpc and within each cone, each point is a group and all points of the same colour belong to the same filament. As $b$ tends to higher and higher values, all groups will be clustered into one massive superfilament, which is unrealistic given that we expect structural collapse in large scale structure to stabilise at scales less than 15 $h^{-1}$ Mpc (Chris Power, private communication). Conversely, if the linking length is too small, prominent superstructures are broken up into several short sub-structures. Additionally, as $b$ drops the total number of groups included in filaments also drops. Therefore at $b = 1$ $h^{-1}$ Mpc we are effectively sampling the distribution of group-group pairs that lie within $1$ $h^{-1}$ Mpc of each other.

\begin{figure}
	\centering
	\includegraphics[width=0.5\textwidth]{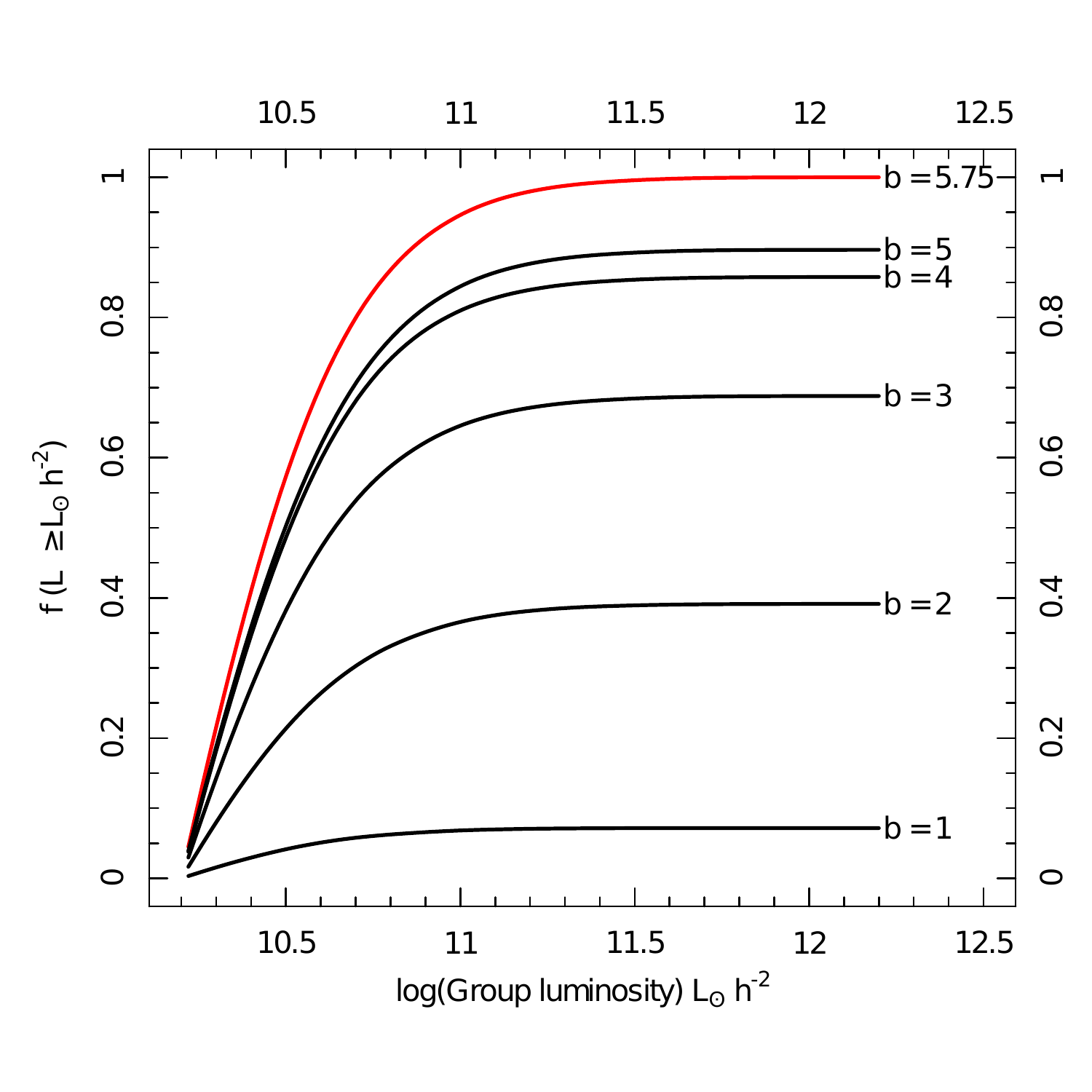}
	\caption{Cumulative fraction of groups in filaments as a function of their total r-band group luminosity, shown for different maximum edge lengths in the minimal spanning tree (given by $b$). As $b$ decreases we begin to only construct filaments between pairs of groups that are in extreme close proximity, and the fraction of high mass groups in filaments drops to 0. Naturally if we raised $b$ to a much higher value all groups would be in a single giant filament. We therefore select the minimum value for $b$ at which $90\%$ or more of galaxies with $L_{\mathrm{Group}} \geq 10^{11} L_{\odot} h^{-2}$ are in filaments; or in other words, $f(L_{\mathrm{Group}} \geq 10^{11} L_{\odot} h^{-2}) \geq 90\%$.}
	\label{fig:fluxbins}
\end{figure}

\begin{figure}
	\centering
	\includegraphics[width=0.45\textwidth]{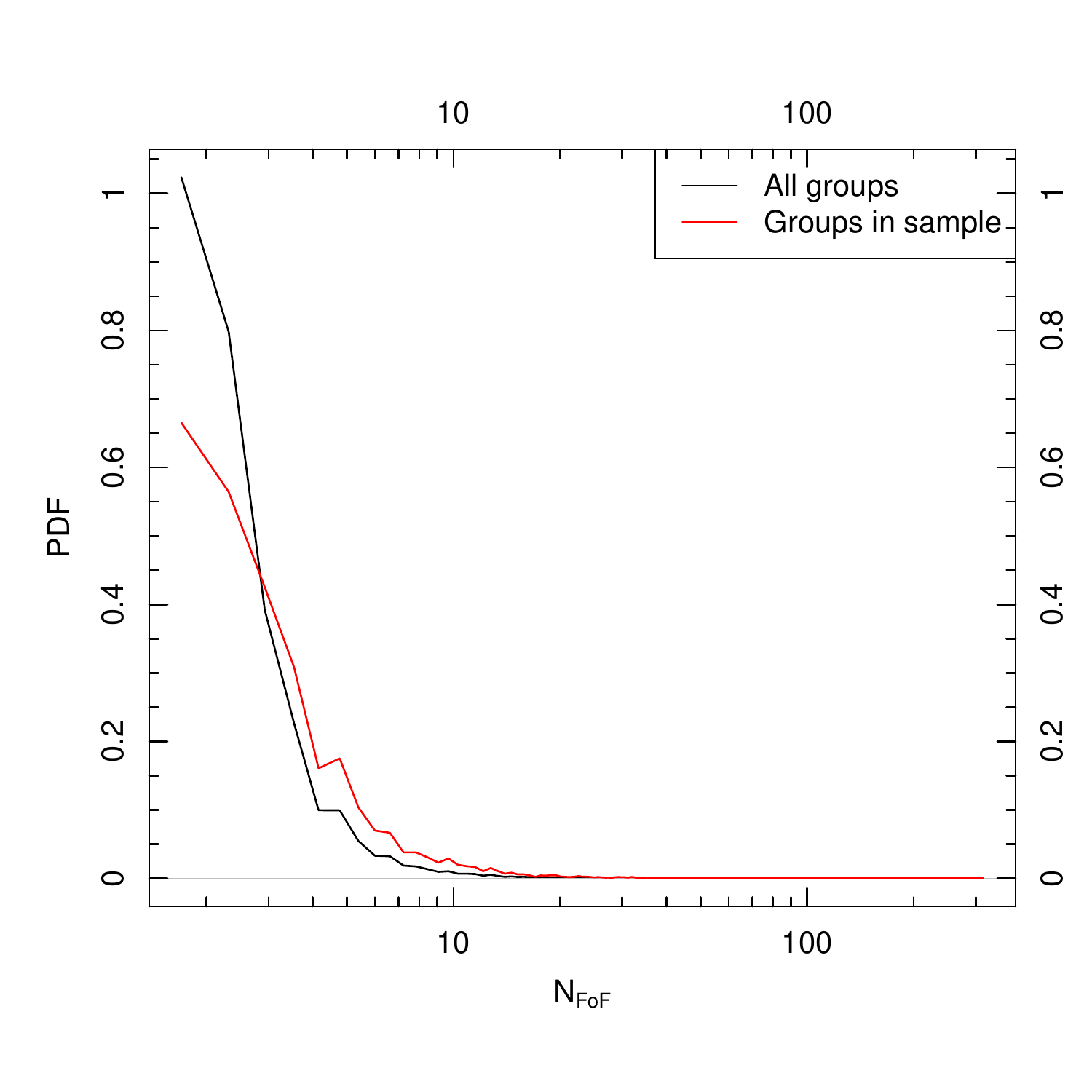}
	\caption{Probability distribution functions of group multiplicity (the number of galaxies per group) shown for the full GAMA Group Catalogue (in black) and for groups in our sample (in red).}
	\label{fig:grouphist}
\end{figure}

To make our selection of $b$ as objective and unbiased as possible, the largest, brightest groups should belong to filaments. We formalise this by defining a bright group as one with $L_{\mathrm{Group}} \geq 10^{11} L_{\odot} h^{-2}$, where $L_{\mathrm{Group}}$ is given by the total group luminosity given in the G$^3$C. This value is roughly equivalent to the $98.65 \%$ quantile in the total range of group luminosities in the sample used. We require the fraction of these bright groups in filaments to be $\geq 0.9$. In Figure \ref{fig:fluxbins} we show the fraction of groups in filaments as a function of $\log(L_{group} / L_{\odot} h^{-2})$ for a different set of values for $b$. This defines $b = 5.75 h^{-1}$ Mpc as this is the minimum length at which this condition is fulfilled, so we trim any edges longer than this value. The MST and filaments shown in Figure \ref{fig:flowchart} are constructed with $b = 5.75 h^{-1}$ Mpc. As expected, as $b$ increases, more groups are linked to the same filament, finally leading to a single massive superstructure, but this would be unphysical. The multiplicity distribution of groups in filaments with $b = 5.75 h^{-1}$ Mpc compared to the full GAMA Group Catalogue is shown in Figure \ref{fig:grouphist}.

\subsubsection{Filament morphology}

Going from a series of links that groups together some points into a common structure, to an understanding of the shape and morphology of that structure is non-trivial. One must define where the edges of the structure are (it may be possible, for example, for a node to exist geometrically near other edges and nodes, but be a dead end itself) as well as the most central part of the filament. To this end, we have developed an algorithm to analyse the structure of a filament called \texttt{walk}. The purpose of this algorithm is to step (or `walk') through the filament and record, for each node, the number of steps required to exit the filament from the nearest end; this is referred to as the \emph{count}. A second property that is recorded is the so-called `branch order' of each node; this value represents the number of branches between the node and the nearest end. A detailed example of this process, as well as a step-by-step analysis of how the algorithm `walks' through the filament is shown in Appendix A. Nodes on branches with one end are said to have a branch order of 1, and this value increases with each intersection. The output of \texttt{walk} after going through this process is a simple table that contains, for each node, the count value which represents a distance, in terms of nodes, between that node and the nearest end of the filament, and its branch order. This approach of splitting filaments into individual branches has previously been used by \citet{Colberg2007}.

The output of \texttt{walk} is fed into a secondary function called \texttt{makebranch} along with the original list of links for the filament. In knowing the count and branch order of each node, this function can travel along the branches of the filament from any given starting point, and search `upwards' or `downwards.' This is a setting specified by the user, and the direction refers to searching `up' to find the centre of the filament, or `down' to get to the nearest end. Potentially, therefore, a user could choose to start at the ends of the filament and find the fastest way to the centre, or vice versa. 

An important setting of \texttt{makebranch} allows the user to instruct the function to avoid reusing nodes that have already been visited by the function. This option is important in determining the primary branches of a filament, which we dub its backbone. To do this, we run \texttt{makebranch} without avoiding nodes first to determine all possible paths that lead from the ends to the filament centre. By then rearranging these branches in descending order and rerunning \texttt{makebranch}, this time avoiding visited nodes, and starting from the two biggest branches, it is possible to determine the longest primary filament that starts at an end, travels to the centre, and moves to another end. In this case `longest' can be determined either by number of nodes, or physical distance.

With the backbone and branches for a given filament, it is possible to objectively look at its morphology. The backbone will always represent the most central route through the filament, and branches will always refer to links emerging from the backbone. The backbone therefore serves as a good measure of the overall extent of the filament, while examining the lengths and sizes of branches, as well as their relative abundances, and provides a measure for the `spread' of the filament. For example, a filament with one large backbone and few to no branches is topologically the same as, or similar to, a straight line; while one with a short backbone and many branches can be seen as less linear.

\subsubsection{Galaxies near filaments}

We now have a set of filaments, each of which contains a number of groups. To associate galaxies with filaments, we now look through each filament and travel along all of its branches, identifying all galaxies within an orthogonal distance $r$ from the filament, with a function called \texttt{Scooper}. For each link along the filament, \texttt{Scooper} identifies all galaxies within a locus at a distance $r$ from the vector that describes that link. Note that \texttt{Scooper} considers all galaxies, both in and out of groups. In three dimensions, this locus is a cylinder with hemispheres at each end. If the distance of the galaxy $d \leq r$ then the galaxy is considered to be associated with that filament. Should a galaxy be within $r$ of two different branches, it is associated with the branch it is closest to.

In order to diminish the effects of redshift space distortions, in this step of the process any galaxy (within our sample) belonging to a group within a filament is automatically assigned to that filament and branch. Instead of the distance to the filament, the distance to the iterative group centre is considered. Visually and statistically (when considering measurements of filament properties, as discussed in Section 4), filaments generated without this redshift space correction look indistinguishable from filaments with the correction, as most of the distorsions are removed in R11 with the creation of the groups.

\subsubsection{Tendrils and voids} 

To identify any underlying structure outside or between filaments, we remove all elements belonging to filaments from our data set. These so called `isolated galaxies' are shown in Figure \ref{fig:flowchart} and are themselves used as points for another MST with a maximum edge length $q$ (the choice for $q$ is detailed below). Once again, galaxies that are part of a single uninterrupted chain of links are classified as being in the same tendril, which are structures akin to filaments but on much smaller scales, formed entirely out of galaxies. Characteristically they branch out from filaments and penetrate into voids. Void galaxies are the last remaining galaxies -- those that were rejected from the MST used to identify the tendrils after the edge trimming. Tendrils and voids are shown as the green and red distributions of galaxies in Figure \ref{fig:flowchart}.

To constrain the maximum edge length $q$ we go back to our second assumption: that void galaxies are only clustered on extremely small lengths. In other words, the spatial two point correlation function of void galaxies should show less signal than the two point correlation function of galaxies in and around filaments. 

\begin{figure}
	\centering
	\includegraphics[width=0.5\textwidth]{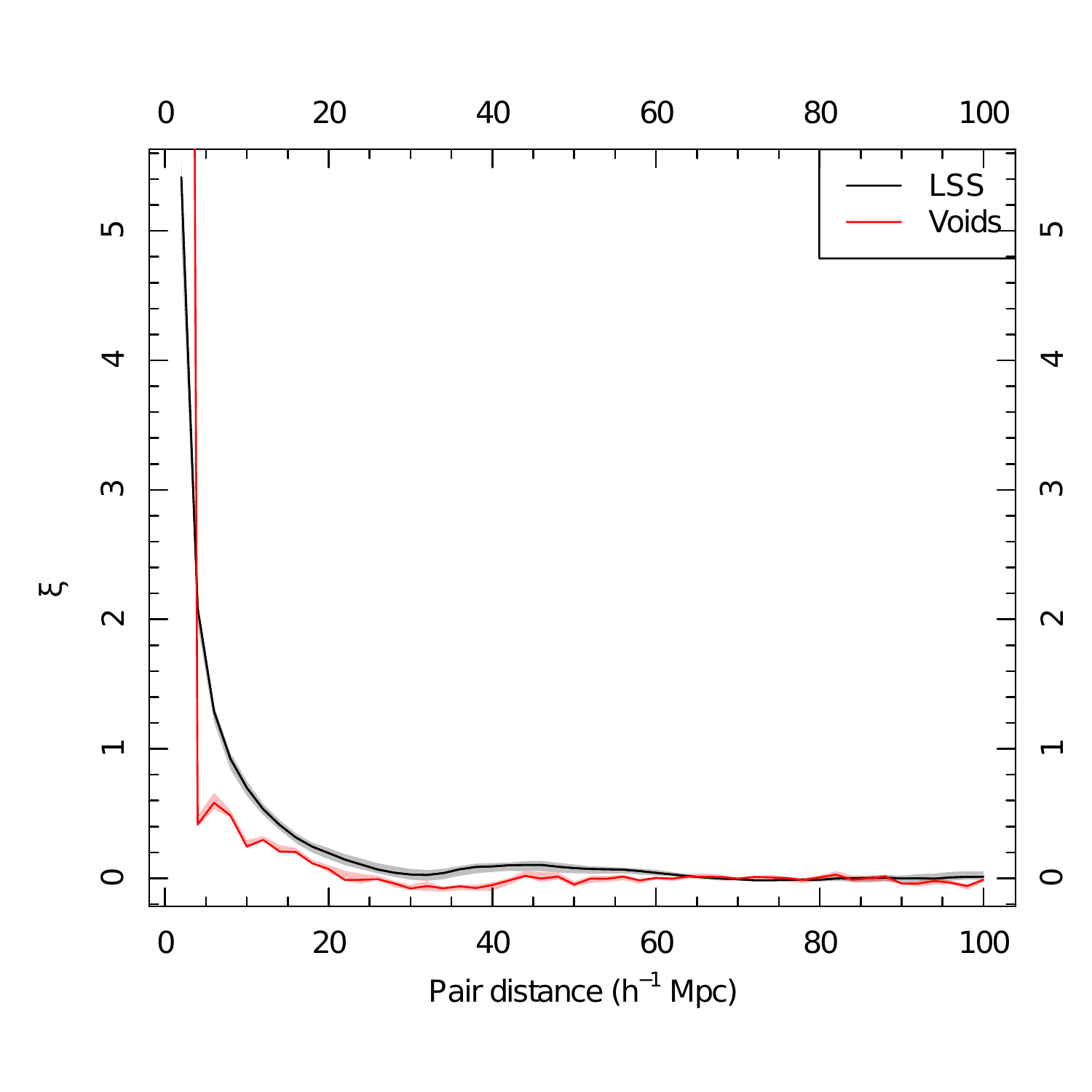}
	\caption{Two point correlation functions as a function of comoving distance for two different galaxy populations. The black line represents galaxies in groups that are in filaments, and within $\sim 4.5$ $h^{-1}$ Mpc of filaments; and the red line shows the function for galaxies in voids. By setting the maximum edge length between filaments to be 5.75 $h^{-1}$ Mpc and roughly 4.5 $h^{-1}$ Mpc between galaxies in tendrils, we ensure that the resulting distribution of void galaxies has no correlation signal. The grey and pink shaded areas show, for each bin, the range of values for two point correlation functions calculated by sub-sampling the GAMA regions, with a jackknife method, and serve as uncertainty estimators.}
	\label{fig:cc}
\end{figure}

The two point correlation function is computed using the estimator from \citet{Landy1993}, namely:

	\begin{equation}
		\xi(r) = \frac{N_r(N_r-1)}{N_d(N_d-1)} \frac{DD(r)}{RR(r)} - \frac{2(N_r-1)}{N_d} \frac{DR(r)}{RR(r)} + 1
	\end{equation}

\noindent where $\xi(r)$ corresponds to the spherically averaged two point correlation function. $DD(r)$, $DR(r)$, and $RR(r)$ refer to the number of pairs separated by a distance $r \pm \ud r$ for data-data pairs, data-random pairs and random-random pairs, and $N_r$ and $N_d$ refer to the number of random points and data points respectively. The random distribution is in the same volume as each GAMA region, and is filled with 100,000 randomly generated points in a spherical pointing; this is to match our volume limited sample. We calculate $\xi(r)$ for each GAMA field separately plot averages. An attempt to estimate uncertainty in the two point correlation functions is made by jackknifing each GAMA region into several sub-regions and recalculating the correlation functions, excluding one subregion at a time. The shaded areas in Figure \ref{fig:cc} around each line show the region occupied by different $\xi(r)$ for each jackknifed region. Because these different samples are so correlated, we caution against strictly interpreting these regions as uncertainties.

We select out all galaxies near filaments and in tendrils and classify them as belonging to large scale structure, and refer to voids as being the remaining population of galaxies. Running the entire algorithm, from generating the filaments to detecting tendrils and voids, takes just over one minute (the most time consuming step is the scooping up of galaxies near filaments). Because of this, we are able to generate a multitude of tendril and void galaxy distributions using different values for $r$ and $q$. The final values for $r$ and $q$ are chosen such that they minimise the integral $\int R^2 \xi(R) \ud R$, with $\xi(R)$ being the correlation function of void galaxies. This is the expression for the volume average correlation function, i.e. $\xi(< R)$. We arrive at $r = 4.13$ $h^{-1}$ Mpc and $q = 4.56$ $h^{-1}$ Mpc. The final parameters used are $b = 5.75 h^{-1}$ Mpc, the trimming length for the MST that identifies filaments, $r = 4.13 h^{-1}$ Mpc, the maximum distance allowed between a galaxy and a filament, and $q = 4.56 h^{-1}$ Mpc, the trimming length for the tendril MST.

We note that our parameter selections are, to some extent, arbitary, just as there are no formal definitions for filaments currently in the literature. The value for $b$ is chosen such that we maximise the number of bright groups included in our filaments, and $r$ and $q$ are chosen such that the void galaxy correlation function is minimised over large distances ($\geq 20 h^{-1}$ Mpc). We may, for example, include groups from the G$^3$C with only one remaining galaxy after making a volume limited sample, but this would lead to a more noisy primary filament MST with many links to small, isolated groups. While it is possible to change and refine the parameter selection process, the overall hierarchy of the large scale structure is very stable with respect to changes in $b$, $r$ and $q$. Varying any parameter by $\pm 1 h^{-1}$ Mpc results in a shift of approximately $5\%$ of galaxies from filaments into tendrils, and tendrils into voids and vice-versa. There is a negligibly small effect on the comparisons to mock filaments discussed later in the paper.



\subsection{Filament catalogue}

\begin{figure*}
	\centering
	\includegraphics[width=0.90\textwidth]{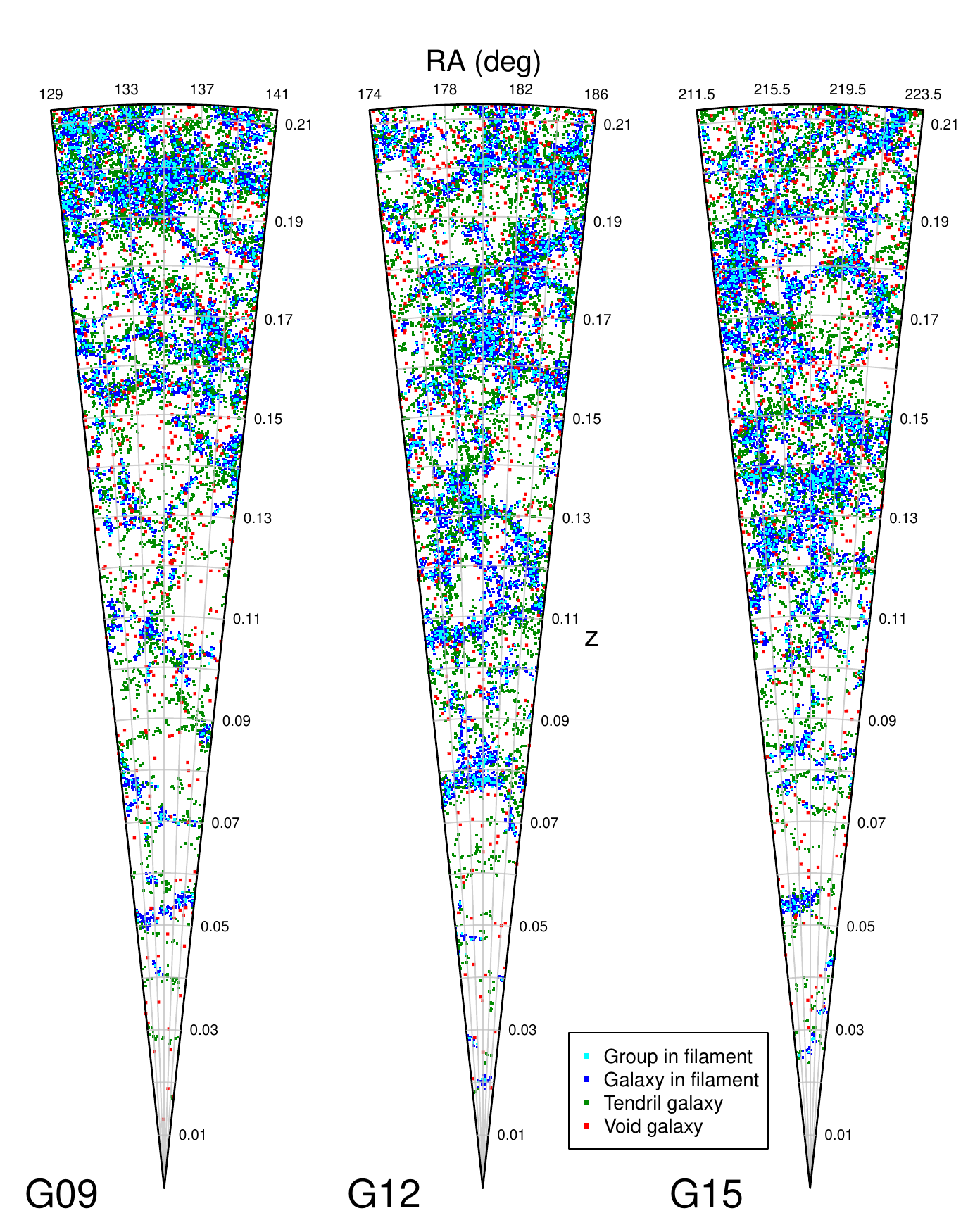}
	\caption{Galaxy distribution in the three GAMA regions, colour coded according to their environment, with groups in filaments, galaxies near filaments, galaxies in tendrils, and void galaxies shown in cyan, blue, green and red respectively. Groups, and galaxies near them form the bulky complexes of large scale structure, tendrils spreading from them in filamentary structures into voids, which seem to be populated by galaxies that appear to be almost uniformly distributed on large scales.}
	\label{fig:lsscones}
\end{figure*}

The algorithm described above is run on all three equatorial GAMA fields, as well as the GAMA mock catalogues. An `overhead' view of all three equatorial GAMA regions, side by side, out to $z = 0.213$ is shown in Figure \ref{fig:lsscones}. Here, cyan points show groups in filaments, blue points correspond to galaxies near filaments, green points to galaxies in tendrils, and red points to void galaxies. It is strikingly easy to visually discern the skeletal pattern traced out by the filaments and their associated groups in blue; these dominate the regions entirely. Tendrils of galaxies appear to be wispy, coherent structures that emerge from dense filamentary regions and either bridge across to other filaments, or terminate within voids. They span a range of different morphologies, as filaments do, with some being very linear (such as the tendril on the upper right region of the top right panel in Figure \ref{fig:lsscones}) while others are more clustered (middle right of the top left panel in Figure \ref{fig:lsscones}). Void galaxies lie in more isolated regions, reinforcing the paradigm that these are unique galaxies that have remained unaffected by their environment for a long period of time, presumably both chemically and dynamically. 


The full catalogue contains just under 650 filaments, with each filament having on average, 8 groups in it. The average length of the backbone of the filament, or in other words, the distance from the most extreme end of the filament to the other, is $\sim 13$ $h^{-1}$ Mpc, while the total of all the links in the filaments is close to $21$ $h^{-1}$ Mpc. Most filaments have 3 branches, and are surrounded by, on average, 46 galaxies. Table \ref{table:summary} summarises these values for all three regions, and for the whole sample. We find the distribution of filaments to be similar between all three regions. On average, we find shorter filaments than those detected in \citet{Tempel2013}.

\begin{table}
	\begin{minipage}{126mm}
	\begin{tabular}{lllllll}
	 &$N_{\mathrm{fil}}$&$\bar{L}_{\mathrm{fil, BB}}$&$\bar{L}_{\mathrm{fil}}$&$\bar{n}_{\mathrm{group}}$&$\bar{n}_{\mathrm{branch}}$&$\bar{n}_{\mathrm{gal}}$\\
	 \hline
	 \textbf{G09}&213&11.6&18.4&7.5&2.8&42.3\\
	 \textbf{G12}&200&14.7&23.6&9.2&3.4&51.6\\
	 \textbf{G15}&230&12.5&20.0&8.0&3.0&44.2\\
	 \hline
	 \hline
	 \textbf{All}&643&12.90&20.7&8.2&3.0&46.0\\

	 \end{tabular}
	 \end{minipage}
	 \caption{Summary statistics of some basic properties of filaments in GAMA. Besides the number of filaments, for each region the following averages are given, the  backbone length, sum of the length of all links, number of groups, branches, and galaxies per filament are given. All lengths are given in units of $h^{-1}$ Mpc. The final row contains these values across all three equatorial fields.}
	 \label{table:summary}
	 
\end{table}

\begin{figure*}
	\centering
	\includegraphics[width=0.85\textwidth]{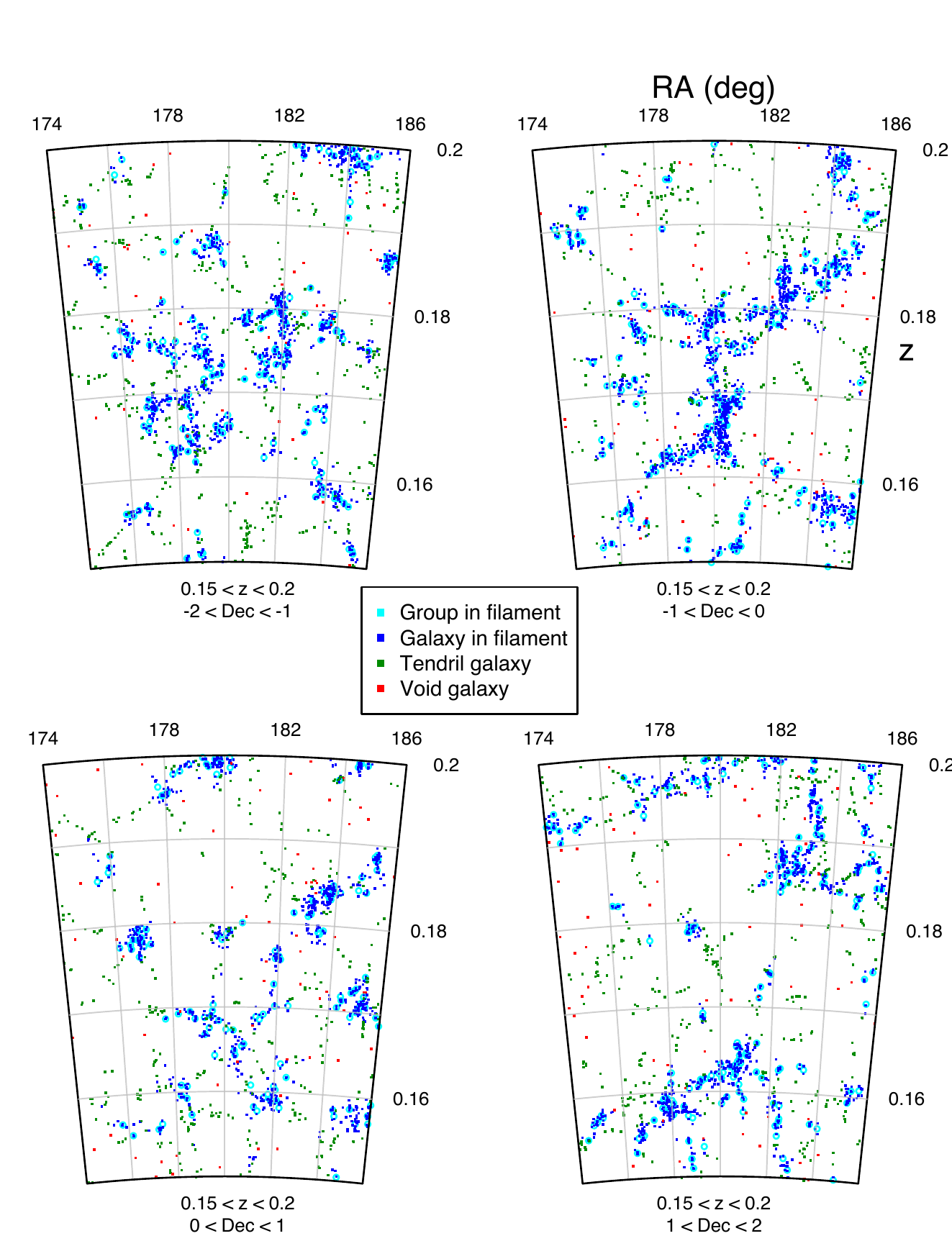}
	\caption{Group and galaxy distribution in G12 colour coded by environment in 4 declination slices. Black circles represent groups in filaments, blue points are galaxies near filaments, green points are galaxies in tendrils, and red crosses are galaxies in voids.}
	\label{fig:g12grid}
\end{figure*}

A more detailed view of a region in the G12 ($174^{\circ} \leq \alpha \leq 186^{\circ}$) field is shown in Figure \ref{fig:g12grid}, with each panel representing a declination slice of $1^{\circ}$, for a redshift range of $0.15 \leq z \leq 0.2$. In this zoomed view it is possible to see the detailed interplay between filaments (blue) and tendrils (green); with the latter branching out from the former and penetrating into voids as coherent structures. It is also possible to see with detail the coherence of structure formed by individual galaxies in tendrils; a notable example is in the top right area of the top right panel of Figure \ref{fig:g12grid}, where a delicate string of galaxies is seen to be curving out of a filament. 

While we make no attempt to identify actual voids in this catalogue, we are able to accurately recover galaxies within voids; these objects can be considered to be extremely isolated with regards to their environment, and can be considered as a separate population of galaxies. The distribution of void galaxies exhibits no inherent structure, although it must be noted that this is in part due to design, as we have selected parameters for our filament finder that produce such a result. An analysis of the structural properties of tendrils, as well as their impact on void sizes will be discussed in Alpaslan et al. (in prep).

We can begin to derive some global properties of galaxies in different environments using existing data catalogues in GAMA. The \texttt{StellarMassesv08} catalogue \citep{Taylor2011} provides stellar mass estimates for galaxies in the three equatorial GAMA regions with $m_r < 19.4$ mag using stellar population synthesis modelling. These models are fit to SEDs built with \emph{ugriz} photometry, obtained via reprocessed SDSS-DR7 \citep{Abazajian2009} imaging frames, provided by the GAMA aperture matched photometry catalogue (\citealp{Hill2011}, Liske et al., in prep). We identify all galaxies in filament backbones, second order branches, third order branches, and tendrils and sum the stellar mass contained within these galaxies in order to obtain an estimate for the fractional distribution of stellar mass amongst these different environments. Note that in order to be mass-complete as well as magnitude complete, we impose a further $\log M_*/M_{\odot} \geq 10.61$ selection cut on our sample. This value is obtained by fitting to the upper 95th percentile of the $\log M_*$ distribution for galaxies with $0.213 \leq z \leq 0.25$ and $M_r$ within 0.05 mag of the sample $M_r$ limit of -19.77 mag. In this sub-sample containing 36\% of galaxies, the mass distribution is as follows: 72.63\% of stellar mass above $\log M_*/M_{\odot} \geq 10.61$ is contained in filaments (39.5\% for backbone galaxies, 23.8\% for second order branch galaxies, 8.10\% for third order branch galaxies and the remaining 1.23\% in further order branches), 23.9\% for tendril galaxies, and 3.42\% for void galaxies. All uncertainties on these percentages are of the order of $\sim 0.05\%$ and are estimated in \citet{Taylor2011} using photometric errors from the GAMA matched aperture photometry catalogue. Note that \citet{Taylor2011} assumes $h = 0.7$.
 
\section{Comparison to mocks}

We now concentrate on comparing the overall properties of filaments in the observed GAMA data, to filaments generated from the mock galaxy and group catalogues. For both data sets, we use the exact same algorithm with identical parameters $b$, $r$ and $q$ and the same sample selection process as described in Section 3, and generate the same hierarchy of catalogues, producing a total of 10 sets of large scale structure catalogues for each region. 

It is important to note that while we are making comparisons between real and mock filaments, these comparisons can only apply to filaments found in the GAMA mocks; moreover, these comparisons are applied on the basis that the algorithm is run on both data sets using the same set of parameters to ensure consistency. Given that the GAMA mocks successfully replicate the number density and luminosity function of the observed GAMA data, and that $b$, $r$, and $q$ depend most strongly on galaxy number density and luminosity, it is acceptable to use the same values for these parameters across both data sets. To ensure that these assumptions are valid, we derive values of $b$, $r$, and $q$ for each individual mock catalogue, and arrive at the following values: $b = 5.75 \pm 0.2$, $r =  4.3 \pm 0.8$, $q = 3.9 \pm 0.7$ $h^{-1}$ Mpc with errors of $1\sigma$ about the mean. The parameters all agree within their errors with values derived for the observed data set.

\begin{figure*}
	\centering
	\includegraphics[width=0.75\textwidth]{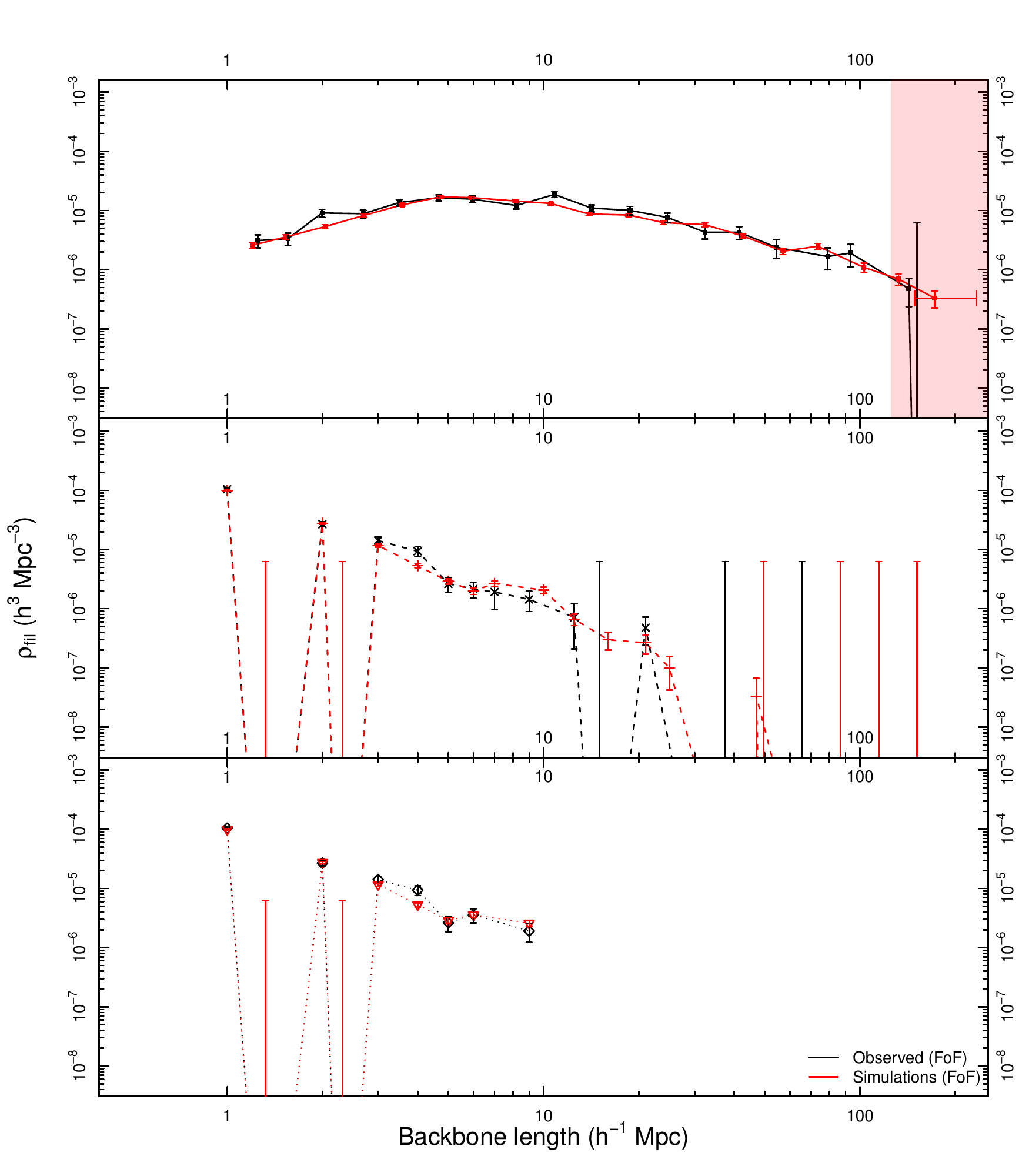}
	\caption{The binned distribution of the number density of filaments as a function of length for various components of filaments with Poisson errors, ranging from backbones (solid lines, top panel) through to branches of order $n = 2$ and $3$ (dashed and dotted lines, in the middle and lower panels respectively). Black and red lines correspond to filaments from the data and mock regions respectively. The $x$-axis positions of the points are the median values within that bin. The horizontal error bar on the final point in the red line shows the $1\sigma$ spread of the fifteen largest mock filaments across all regions and volumes. Bins with no detections show only an upper limit derived from Poisson statistics. The shaded region marks distances at which the geometry of the GAMA regions means the backbone lengths are poorly constrained.}
	\label{fig:b75alldist}
\end{figure*}

A comparison of filament lengths is shown in Figure \ref{fig:b75alldist} which shows, for observed and mock data (the black and red lines respectively) the binned abundance distribution of the lengths of filament components as a function of the number density of filaments. The backbone length is a good indicator of the overall span of a filament across its dominant axis as it traces the longest possible path from one end of the filament to the other through its central node; these are shown by the solid lines in Figure \ref{fig:b75alldist}, with vertical error bars giving $1\sigma$ uncertainty ranges. The horizontal error bar on the final point for the mock filaments on the right of the top panel marks the $1\sigma$ distribution in maximum filament lengths across mock filaments. For bins with only one object, we calculate upper error limits based on Poisson statistics. Some filaments in the GAMA mock catalogues are longer than those observed in the real data, but sample variance does not allow us to draw any conclusions on this. To first order there is a remarkably good agreement between the mock and observed filaments across 2 orders of magnitude of scale. Similarly, tendril lengths agree very well between simulated and observed data.

\begin{figure*}
	\centering
	\includegraphics[width=0.4985\textwidth]{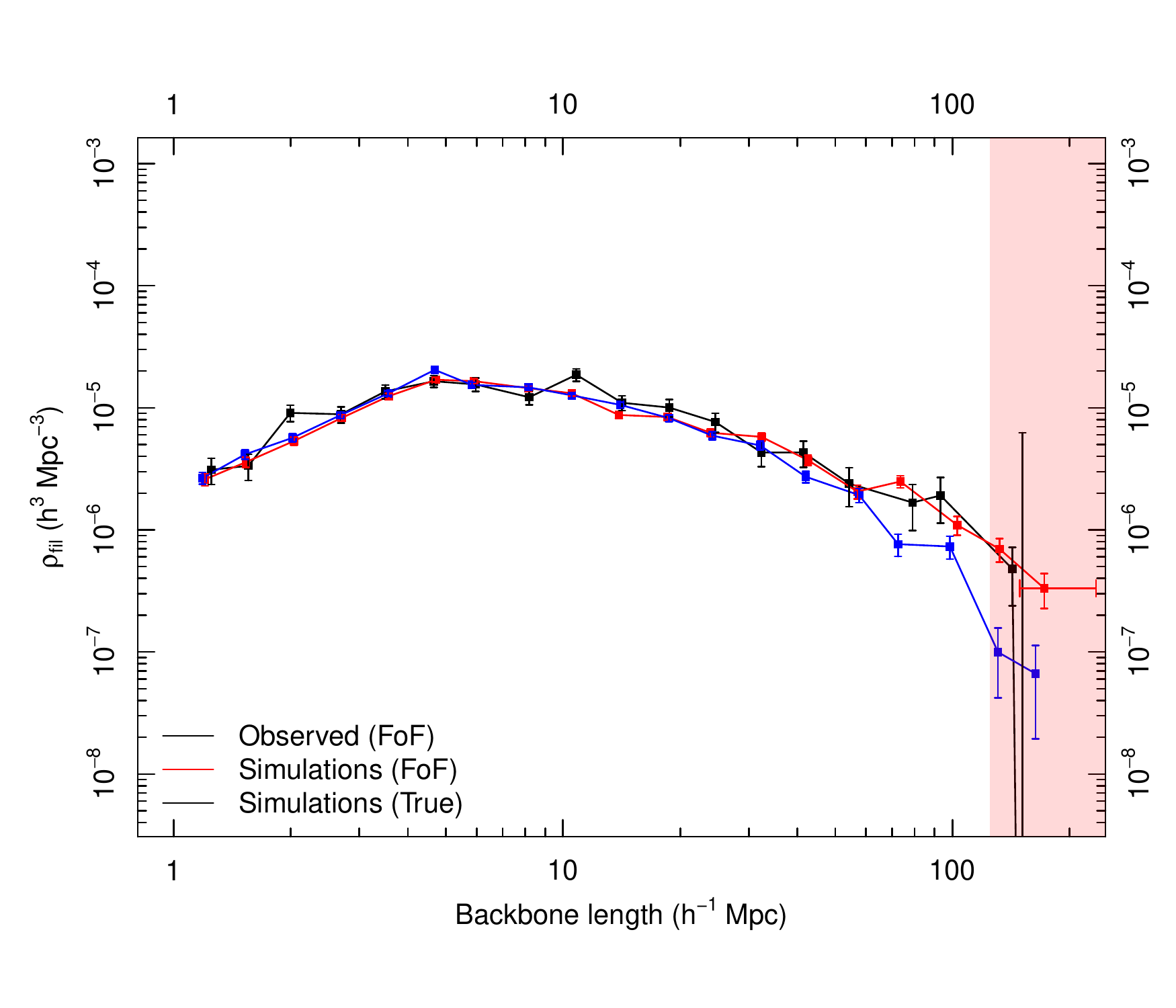} \includegraphics[width=0.4875\textwidth]{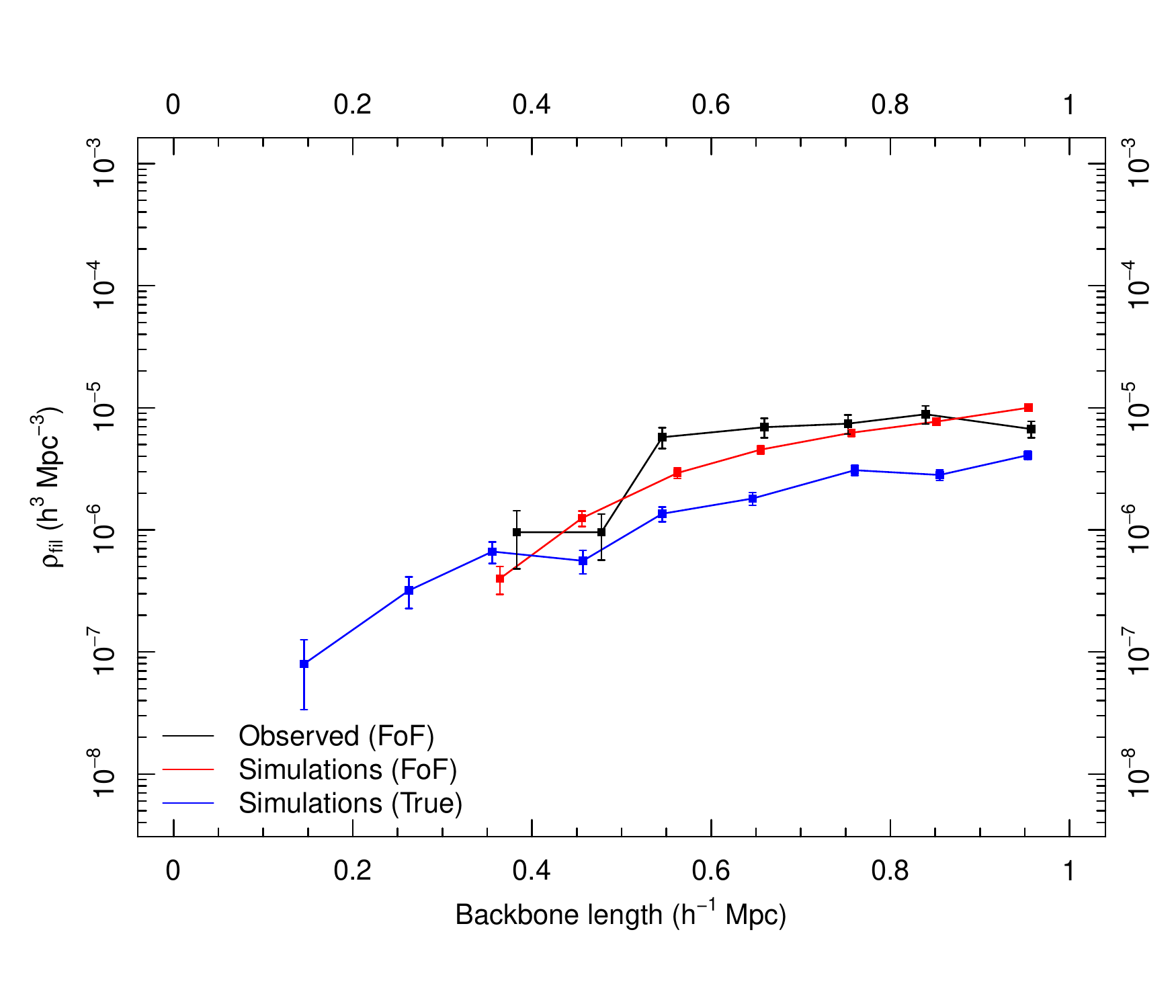}
	\caption{\textit{Left:} In the same manner to Figure \ref{fig:b75alldist}, the number density of filaments as a function of backbone length is shown. The black, red and blue lines each correspond to filaments in observed groups, groups recovered from simulations using the R11 groupfinder, and the intrinsically known groups from the simulations. \textit{Right:} The distribution of backbone lengths for filaments whose edges cannot exceed 1 $h^{-1}$ Mpc. These filaments are all group-group pairs. Notably, we see more short filaments constructed from the intrinsic groups. We observe that there are more short filaments of intrinsic groups compared to the observed and FoF groups.}
	\label{fig:halos}
\end{figure*}

Given that the G$^3$C has been generated by calibrating the FoF algorithm against mock galaxies whose intrinsic grouping is known, it is an interesting exercise to generate filaments using haloes instead of groups. We again apply the same algorithm, with the same sample selection (we select haloes instead of groups of galaxies) and generate filaments of haloes, whose backbone length is shown on the left panel in Figure \ref{fig:halos} in blue. Using the same values for $b$, $r$, and $q$ in this case ensures that any difference in results for these filaments will be due to how the groupfinder in R11 breaks haloes apart into groups. These can be considered the `true' mock filaments, as they are not subject to biases in the FoF algorithm. Halo filaments are remarkably similar to mock and observed filaments, as shown in the left panel of Figure \ref{fig:halos}. We expect FoF filaments to be longer than halo filaments, however, as the FoF algorithm will occasionally break a halo into multiple groups; this effectively means that the MST has an extra stepping stone between two haloes and is therefore able to form structures with shorter links that are less likely to be trimmed later. There is an equal chance for the groupfinding algorithm to merge multiple haloes into a single group, depending on the halo mass range being considered.

Similarly by reducing the maximum edge length to 1 $h^{-1}$ Mpc, we begin to examine the group-group pairs that are within 1 $h^{-1}$ Mpc of each other; all the FoF filaments in this sample consist of two neighbouring groups (most halo filaments are composed of 2 or 3 haloes). We can see in the right panel of Figure \ref{fig:halos} that for intrinsic group filaments there exist shorter group-pair filaments compared to the observed data.

\begin{figure}
	\centering
	\includegraphics[width=0.5\textwidth]{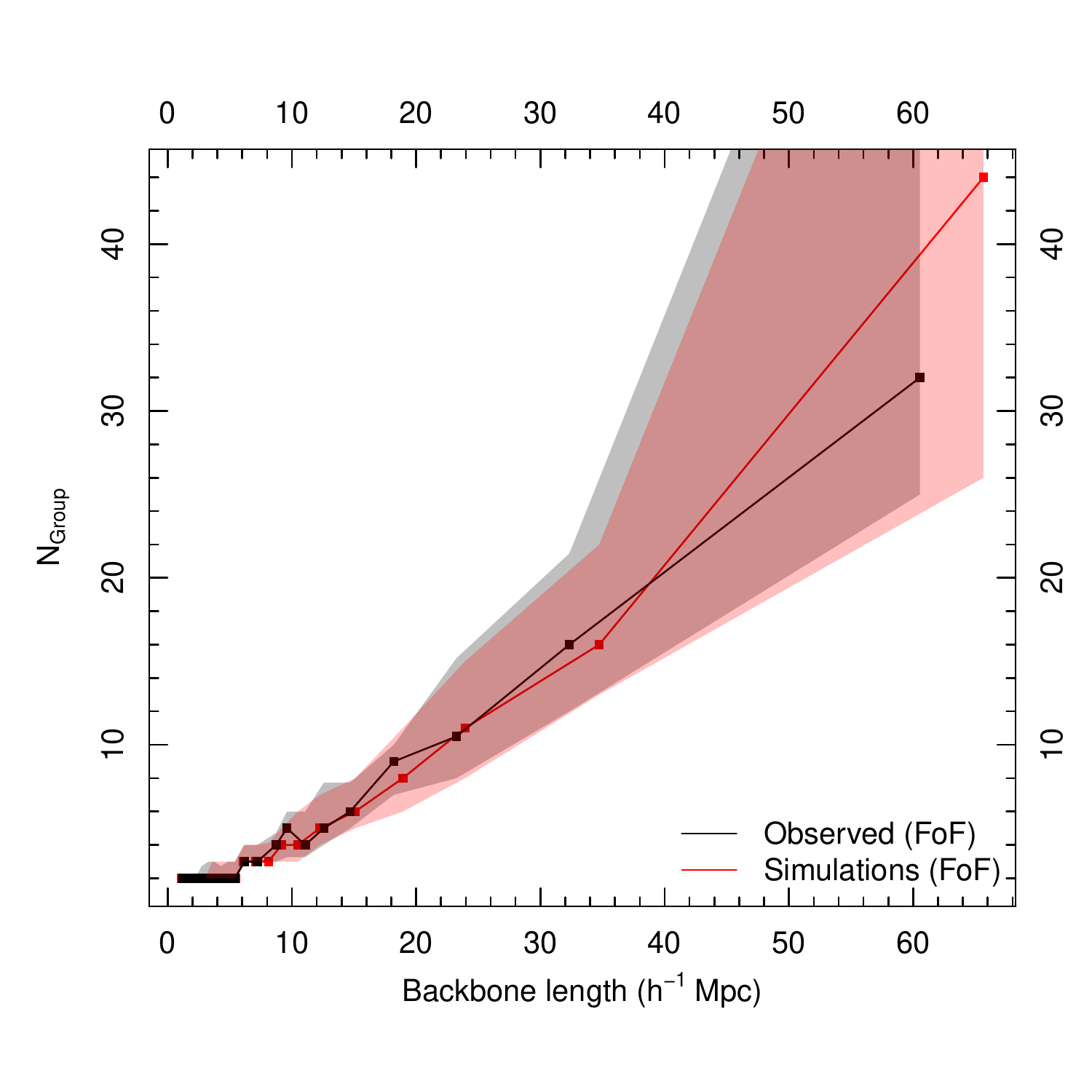}
	\caption{The relationship between backbone length and the number of groups in the backbone in bins containing equal numbers of filaments for observed data and FoF mock groups, shown in black and red respectively. The shaded regions denote $1\sigma$ intervals around the mean; points with no shaded region around them are single entries. This data is binned along the $x$-axis, and there are bins where there is no data; in these cases the point is omitted.}
	\label{fig:b75sizedist}
\end{figure}

Both data sets have filaments that grow larger in a similar way as a function of number of groups. In Figure \ref{fig:b75sizedist} we show the length of the backbone of the filament as a function of the number of groups in the backbone of the filament, for observed and mock data in black and red respectively. The shaded regions about each point show $1\sigma$ spreads about the mean for the filaments in that length bin, and binning is made so that each contains 20 filaments. The growth of filaments is very similar between observed and mock data, with no statistically significant differences.

\begin{figure}
	\centering
	\includegraphics[width=0.5\textwidth]{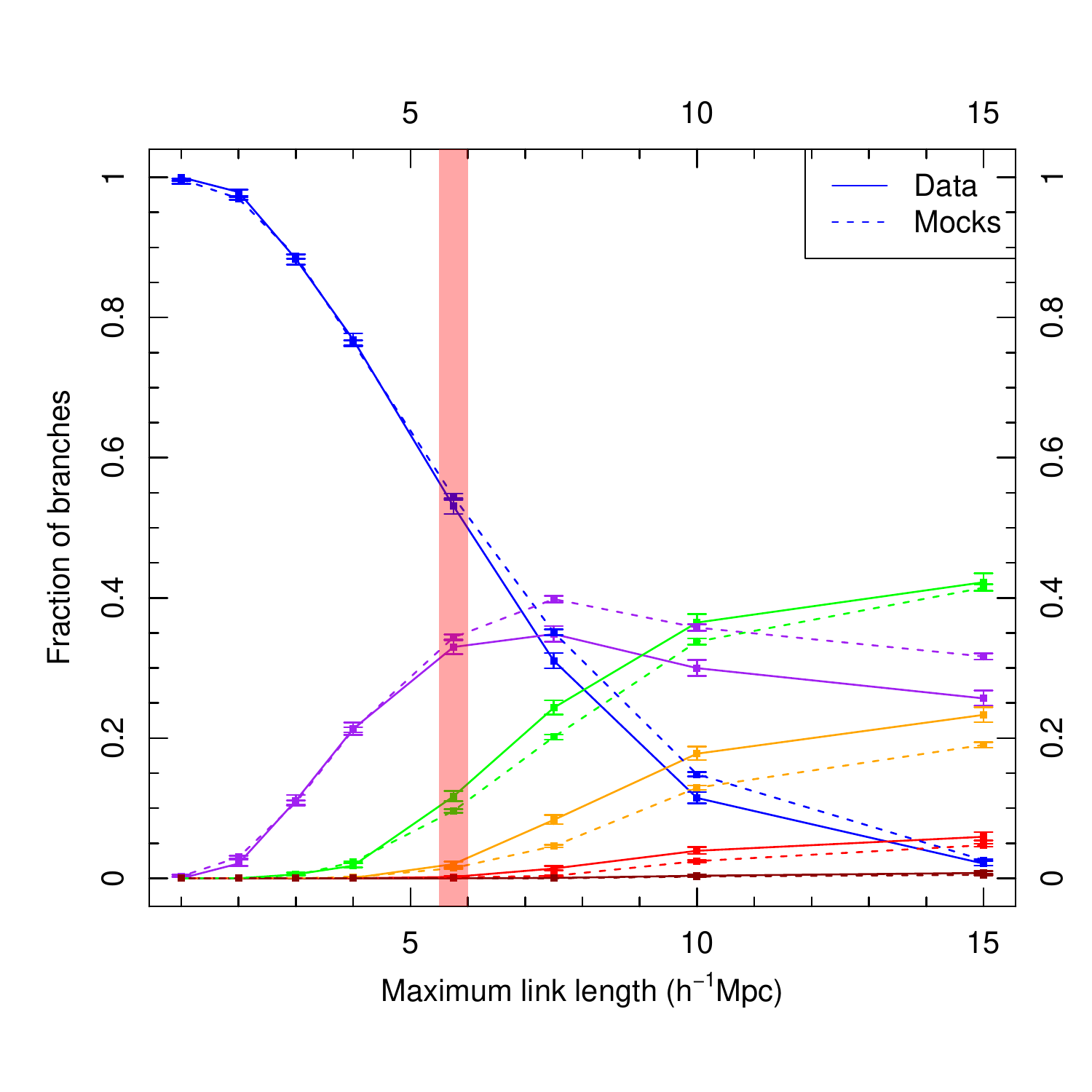}
	\caption{Comparing the `complexity' of filaments in observed data and simulations; where by complexity we refer to the fraction of branches that are backbones, and higher order branches. The solid lines represent data, and the dashed lines, mocks; as the change in colour, they show increasing branch orders, from first to fifth order shown in blue, purple, green, orange and red. As $b$ increases, more complex filaments with more higher order branches are formed. At $b = 5.75 h^{-1}$ Mpc, shown by the shaded region, observed and mock filaments have similar fractions of branches, aside from a slight overabundance of third order branches in observed filaments. The error bars show $1\sigma$ uncertainties about the population fraction.}
	\label{fig:complexity}
\end{figure}

In a similar way, we can examine the complexity of filaments as a function of the maximum linking length allowed between groups during the MST process. Here, complexity refers to the relative fractions of branches of different order; a `simple' filament being one with only a backbone (so only $n = 1$ branches), and a `complex' filament one where there are many orders of branches (branches with $n > 1$). As the maximum linking length in the MST tends towards smaller values, the complexity of filaments decreases, as trees are only allowed to exist between very close group neighbours, and these tend to be simple group-group pairs, as with the population of filaments shown in Figure \ref{fig:halos}. In Figure \ref{fig:complexity}, we show the relative fractions of branches within filaments as a function of maximum linking length $b$. As in Figure \ref{fig:difftrees}, these filaments are all constructed from the sample of galaxies and groups, with only the MST parameter $b$ varying. The solid and dashed lines show the fractions of branch orders for observed and mock data respectively; and the colour of the line represents the branch order. Blue, purple, green, orange and red show the fraction of branches of order $n = 1~,n=2~,n=3~,n=4~,\rm{and}~n=5$ in all filaments for that particular value of $b$. The errors for each point confidence estimates on population proportions derived from the Beta function, as described in \citet{Cameron2011}. 

The points in the shaded region of Figure \ref{fig:complexity} show the relative branch fractions for the same filaments displayed in Figures \ref{fig:b75alldist} and \ref{fig:halos}. The relative fraction of third order branches is slightly higher for observed filaments, otherwise both sets of filaments are very similar in their morphology. The difference between the two populations decreases sharply at lower values of $b$ and for $b \leq 5 h^{-1}$ Mpc there is no difference between the data and the mocks, however at $b = 15 h^{-1}$ Mpc there is a more notable difference between the fractions of some components, most notably second and third order branches.

\begin{figure*}
	\centering
	\includegraphics[width=0.85\textwidth]{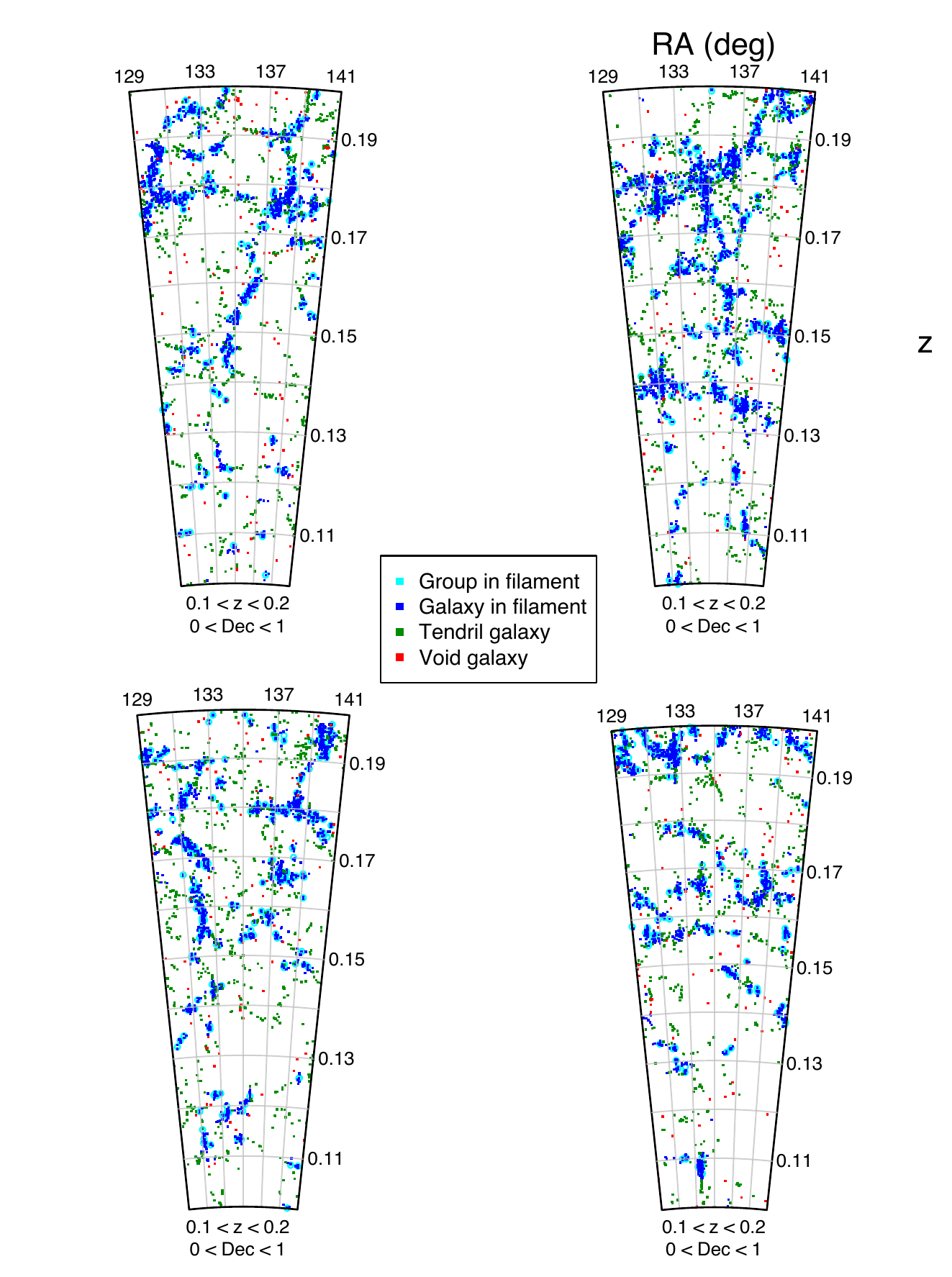}
	\caption{The same region in G09 with different galaxy populations (colour coded as in Figure \ref{fig:g12grid}), with observed data shown in the bottom right and the other panels consisting of mock data. The similarity between all four fields is apparent, and serves to visually highlight the success of the mock catalogues in reproducing large scale structure.}
	\label{fig:m09grid}
\end{figure*}

Beyond the power of the two-point correlation function, this analysis confirms that the GAMA mocks successfuly reproduced the observed distribution of galaxies on large scales. It is very difficult to visually distinguish between real and mock data when looking at large scale structure maps, as shown in Figure \ref{fig:m09grid}. Once again, figures \ref{fig:halos} and \ref{fig:b75sizedist} show the overall similarity between observed and mock filaments.

\section{Discussion and summary}

We have presented a method to systematically identify and categorise large scale structures in the Universe, as well as identify different populations of galaxies in different density environments. Our algorithm is based on using minimal spanning trees to identify filaments composed by groups, around which we identify nearby galaxies that are associated with each filament. The remaining population of galaxies is then classified as tendril galaxies or void galaxies using a second minimal spanning tree. The parameters we use for this approach are selected by optimising for a large scale structure that obeys two assumptions: that the brightest groups be in filaments, and that the distribution of void galaxies show much less structure than for filaments and clusters. We are able to generate large scale structure catalogues for the three equatorial GAMA fields, as well as 9 mock galaxy volumes for each field, adding to a total of 3 observed LSS catalogues and 27 mock LSS catalogues.

Overall, mock large scale structure strongly resembles observed large scale structure and is virtually indistinguishable by eye (as shown in Figures \ref{fig:complexity} and \ref{fig:m09grid}). In our filament analysis we are able to decode the topology of filaments into a primary backbone of links that travels from one end of the filament to the other across its centre, and various tributary branches that connect up to this central spine. We show in Figure \ref{fig:complexity} that filaments in simulations have complexities that match very closely with observed filaments, for the value of $b$ that we have used.

We also identify a secondary population of galaxies that lie in smaller, but still coherent structures which we refer to as `tendrils' of galaxies. Tendrils are much shorter (10 $h^{-1}$ Mpc on average) than filaments and contain fewer galaxies than filaments contain groups; and have much simpler morphologies. Visually they appear to form bridges between filaments and, perhaps more crucially, jut out into voids, and in some cases, even bisect voids. This will be discussed in more detail in Alpaslan et al. (2013). 

With GAMA we benefit from an extremely complete survey that has a very high target density, revealing that there is far more underlying structure behind the brightest galaxies and groups that form the skeletal signatures of filaments and large scale structure. As we revisit the same patch of sky and conduct deeper, more complete observations, we find that filaments span larger widths, leading to voids becoming smaller. It is therefore somewhat of a misnomer to still refer to these structures as filaments as they are more complex than simple one dimensional structures; but it is also a valid statement to claim that as we are able to probe deeper into the Universe and conduct wide surveys at lower magnitude depths, our understanding of large scale structure is bound to change. We look forward to results from future galaxy surveys to further illuminate this subject.

\section*{Acknowledgements}

The authors would like to thank the referee for their kind and constructive comments on this work. MA would like to acknowledge funding from the University of St Andrews and the International Centre for Radio Astronomy Research. ASGR is supported by funding from a UWA Fellowship. PN acknowledges the support of the Royal Society through the award of a University Research Fellowship and the European Research Council, through receipt of a Starting Grant (DEGAS-259586). MJIB acknowledges the financial support of the Australian Research Council Future Fellowship 100100280. TMR acknowledges support from a European Research Council Starting Grant (DEGAS-259586).

GAMA is a joint European-Australasian project based around a spectroscopic campaign using the Anglo-Australian Telescope. The GAMA input catalogue is based on data taken from the Sloan Digital Sky Survey and the UKIRT Infrared Deep Sky Survey. Complementary imaging of the GAMA regions is being obtained by a number of independent survey programs including GALEX MIS, VST KIDS, VISTA VIKING, WISE, Herschel-ATLAS, GMRT and ASKAP providing UV to radio coverage. GAMA is funded by the STFC (UK), the ARC (Australia), the AAO, and the participating institutions. The GAMA website is http://www.gama-survey.org/.

\appendix
\section{Filament walker and finding backbones}

In this appendix we give a brief explanation of the algorithm we use to step through the filaments we created using minimal spanning trees. The purpose of these functions is to be able to systematically determine the internal structure of a filament; that is to say, where its edges are, where the most dense nodes of the filament are, and how to travel from one region of the filament to another in the most efficient path possible.

Throughout of this appendix, we adopt the following nomenclature, borrowed from graph theory: a \emph{graph} represents a collection of points, or \emph{nodes} and \emph{edges} are defined as the lines which connect them. In this work, our graphs represent filaments, with nodes corresponding to groups of galaxies whose positions are defined by their median RA, Dec and redshift. We define \emph{branches} as individual links of nodes that form part of the full tree, and \emph{ends} as any nodes that have only one edge.

\begin{figure*}
	\centering
	\includegraphics[width=1\textwidth]{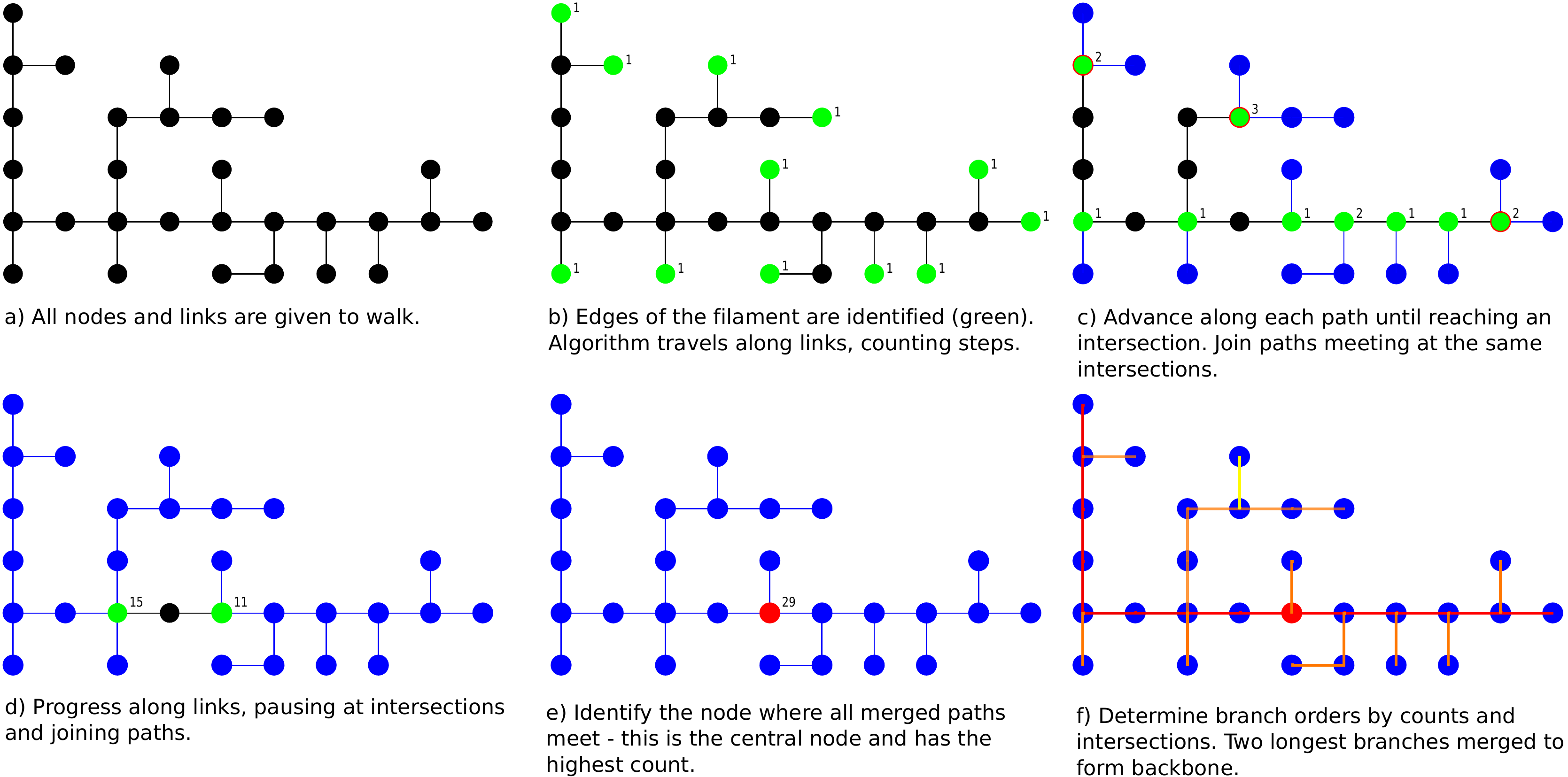}
	\caption{An overview of the process by which the topology of a filament is determined. The top left panel shows all nodes and links for an example filament (circles and lines respectively). In each panel, green objects represent where the algorithm is, while blue ones represent visited objects and black ones, unvisited objects. From here, \texttt{walk} identifies all the ends of the filament (shown in panel b) and travels along them, stopping at intersections and merging all paths that reach the same intersection (panels c and d). The algorithm associates a count value to each node (shown for the green nodes on the top right in each panel), which is the number of steps required to reach the end of the filament. The count of the centre of the filament at an intersection, the counts along each branch are summed up and assigned as the count value for that node. Therefore, the node at which all branches meet will have the highest count, and be determined to be the centre of the filament, as shown by the red node in panel e. The output of \texttt{walk} is fed into \texttt{makebranch}, which analyses this output and uses it to construct branches for the filament, and assign orders to them. These are shown in panel f, with first, second and third order filaments shown in red, orange and yellow respectively. The backbone is then defined as the single path that travels along the two first order branches.}
	\label{fig:algorithm}
\end{figure*}

All algorithms have been written using R. To illustrate the function of all of our algorithms, we will apply them to a sample filament, which is shown on the top right panel of Figure \ref{fig:algorithm}. In this schematic (collapsed into two dimensions from a sample 3D filament), each circle point marks a galaxy group (or node) and lines represent links between nodes.

\subsection{walk}

The most important algorithm in analysing the structure of a filament is \texttt{walk}. This algorithm's purpose is to start at the ends of the filament (defined as being nodes that have only one edge) and travel along all the links in the filament until they have all been visited. For each node, walk assigns a \textit{count}, which is effectively the number of steps required to get from that node to the nearest filament end. The algorithm also keeps track of the \textit{order} of each node. Nodes at the filament ends are said to be of order 1, and this increases each time the algorithm goes past an intersection. In the example filament shown in Figure \ref{fig:algorithm}, the nodes shown in green in panel b have orders of 1, while nodes shown in green in panel c have orders of 2. One important note to make is that the count and branch order given for a particular node are not with respect to the nearest end, but are a sum of all branches leading out from it. 

Briefly, the \texttt{walk} algorithm works as follows:
\begin{itemize}
	\item{To begin with, \texttt{walk} identifies all nodes in the tree that have only one edge. This implies that they lie at the ends of branches. In the example filament, these are the green nodes in panel b.}
	\item{The algorithm then goes through each end node and progresses along that path until it reaches an intersection. In the case of this example filament, the first intersections are the green nodes in panel c, with the blue nodes and links representing the path the algorithm has taken to reach those nodes.}
	\item{If an intersection has been reached by more than one branch, these are then merged, before the algorithm continues to progress along the edges leading from them. Intersections that have only been reached by one edge do not progress. In panel c, \texttt{walk} will only continue walking from the green nodes circled in red, as these have two links leading out of them.}
	\item{This pattern of stopping at intersections and merging branches continues until all nodes have been visited. In panel d, the process is shown for the step at the intersection where the final two pathways are about to meet - while these nodes had been reached by \texttt{walk} as early as panel c, they have only now been arrived at from two links, meaning the algorithm can walk along them now. In panel e, all paths have merged at the central node, shown in red.}
\end{itemize}

The output of \texttt{walk} after going through this process is a simple table that contains, for each node, the count value which represents a distance, in terms of nodes, between that node and the end of the filament, and its branch order.

\subsection{makebranch}

We now have, for each node, an understanding of how far away it is from the ends of the filament, and how many intersections it takes to get from that node to the end. We feed this information to an algorithm called \texttt{makebranch} that is capable of starting at any node specified by the user and travel `upwards' or `downwards' along the links in the filament; that is to say, from a given node it can either travel to the neighbouring node with a higher or lower count respectively. Travelling downwards will lead it to the nearest filament end, while traveling upwards will lead it to the filament centre. 

To determine the location of the filament's backbone, we first run \texttt{makebranch} by starting it on the ends of the filament and tell it to travel upwards; this gives all possible paths to the filament centre. We rearrange these branches in descending order and have \texttt{makebranch} to travel along them again, this time instructing it to avoid revisiting nodes it has already been to. By starting at the biggest branch and travelling to the centre, then doing the same for the second biggest branch and so on, this algorithm is able to determine the longest path that goes from one end of the filament to the other while travelling through the central node. These are referred to as the first order branches; any path that branches off from these are second order branches, and so on. In other words, a path that intersects with a branch of order $n$ is assigned an order of $n+1$. This implies that there will only be two branches of order 1, but any number of subsequent branches. The higher the number of branches in a filament, the more complicated its morphology.

All the branches in the sample filament are shown in panel f of Figure \ref{fig:algorithm}. First order branches are shown in red; therefore the single unbroken red path is the backbone of the filament. Orange paths are second order branches, and the single yellow path is the third order branch, as it is the only path that intersects with a second order branch.

\section{GAMA Structure Catalogue}

We run the large scale structure algorithm separately on all three equatorial GAMA volumes as well as the GAMA mock cones. The algorithm produces, for each volume, the following catalogues:

\begin{itemize}
	\item \textbf{Filaments}

	This lists all filaments composed of groups of galaxies, giving them a unique identifier in FilID - the first digits of which correspond to the equatorial region the filament is in.  There is also information pertaining to the number of branches the  filament has, as well as the number of groups. The total length of all links in the filament is given, as well as the length of the backbone.

	\item \textbf{FilBranches}

	This catalogue lists all branches present within filaments in Filaments. Each branch is given a unique identifier and the filament it belongs to is identified as well. The order of the branch is given, as well as the number of groups it has, and its length.

	\item \textbf{FilGroups}

	This catalogue contains the groups that are within filaments. They are identified by their GroupID as given in R11's catalogue, as well as their RA, Dec and median redshift. The groups' 3D comoving cartesian coordinates are also provided, as well as the branch they belong to, its order, and the filament they belong to. 

	\item \textbf{FilGals}

	In this catalogue, all galaxies that are within a certain orthogonal distance of filaments are listed. The GAMA CATAID (an internal unique galaxy identifier) for each galaxy is given, along with 3D comoving cartesian coordinates, as well as the orthogonal distance to the nearest branch of a filament, whose IDs are given.

	\item \textbf{FilLinks}

	This simply contains a list of links between groups used to construct the filaments. The groups are identified by their GroupIDs. This catalogue can be used to reconstruct, visually, the links between groups in filaments, but can also be used to identify groups that are `intersections' - that is to say, groups that have 3 or more links to other groups.

	\item \textbf{Tendrils}

	Moving from groups to galaxies, this catalogue is analogous to Filaments in that it contains the top level structures formed by galaxies that are not included in filaments. Each tendril is given a unique ID, and their length and number of galaxies are specified.

	\item \textbf{TendrilGals}

	This is the catalogue of all galaxies in tendrils. Their CATAID is given, as well as their 3D comoving cartesian coordinates, and the ID of the tendril they belong to.

	\item \textbf{TendrilLinks}

	A second list of links, this time for the tendrils. Now, galaxies are identified by their CATAIDs.

	\item \textbf{VoidGals}
	
	Finally, this catalogue lists all galaxies that are not associated with any filaments or tendrils.

\end{itemize}

Each catalogue links with others using a series of unique identifiers for each type of structure: filament, branch, group, and galaxy. Separate catalogues describe tendrils and the galaxies in them in a similar way, and voids are isolated from them all. The links within structures are also given (i.e. the links of the minimal spanning tree after edges are cut), both for filaments and tendrils. All of this allows a user to fully reconstruct the large scale structure of the GAMA regions easily. For example, a user may wish to identify all galaxies associated with the longest filament in G09; this is easily done first by using \textbf{Filaments} to search for the longest filament whose identifier begins with 9, then going to \textbf{FilGals} and selecting all galaxies with a filament ID that matches the filament found.

\footnotesize
\bibliographystyle{mn2e}
\setlength{\bibhang}{2.0em}
\setlength{\labelwidth}{0.0em}
\bibliography{filaments}
\normalsize

\label{lastpage}

\end{document}